\documentclass{mn}[3]
\usepackage{rotating}
\input psfig.sty
\begin{document}

\title[binary evolution \& blue stragglers]
{Mass transfer from a giant star to a main sequence companion and
its contribution to long-orbital-period blue stragglers}
\author[Chen and Han]{Xuefei Chen$^{1}$\thanks{
xuefeichen717@hotmail.com} and Zhanwen Han$^{1}$\\
$^1$National Astronomical Observatories/Yunnan Observatory, CAS,
Kunming, 650011, P.R.China} \maketitle

\begin{abstract}
Binary population synthesis shows that mass transfer from a giant
star to a main-sequence (MS) companion may account for some
observed long-orbital period blue stragglers. However, little
attention {\bf is paid to this blue straggler formation scenario}
as dynamical instability often happens when the mass donor is a
giant star. In this paper, we have studied the critical mass
ratio, $q_{\rm c}$, for dynamically stable mass transfer from a
giant star to a MS companion using detailed evolution
calculations. The results show that a more evolved star is
generally less stable for Roche lobe overflow. Meanwhile, $q_{\rm
c}$ almost linearly increases with the amount of the mass and
angular momentum {\bf lost} during mass transfer, but has little
dependance on stellar wind. To conveniently use the result, we
give a fit of $q_{\rm c}$ as a function of the stellar radius at
the onset of Roche lobe overflow and of the mass transfer
efficiency during the Roche lobe overflow.

To examine the formation of blue stragglers from mass transfer
between giants and MS stars, we have performed Monte Carlo
simulations with various $q_{\rm c}$. {\bf The simulations show
that some binaries with the mass donor on the first giant branch
may contribute to blue stragglers with $q_{\rm c}$ obtained in
this paper but will not from previous $q_{\rm c}$. Meanwhile, from
our $q_{\rm c}$, blue stragglers from the mass transfer between an
AGB star and a MS companion may be more numerous and have a wider
range of orbital periods than those from the other $q_{\rm c}$.}

\end{abstract}


\section{Introduction}
Blue stragglers (BSs) are stars that have remained on the main
sequence for a time exceeding that expected, for their masses,
from standard stellar evolution theory. The existence of BSs
indicates an incomplete understanding of stellar evolution and
also of star formation within clusters \cite{str93}. {\bf Since}
they are bright and blue, these objects may affect the integrated
spectra of their host clusters by contributing excess spectral
energy in the blue and ultraviolet. BSs are therefore important in
studies of population synthesis \cite{xin05}. Much evidence shows
that these strange objects are relevant to primordial binaries
\cite{fer03,dpa04,map04}. Binary coalescence from a contact binary
is a popular hypothesis for single BSs and it is believed that
these contact binaries are mainly from case A mass transfer
\footnote{\bf According to the evolutionary state of the primary
at the onset of mass transfer, three mass transfer cases are
defined, i.e. case A for the primary being on the main sequence,
case B for the primary after the main sequence but before central
He burning, and case C for the primary during or after central He
burning \cite{kip67}.} \cite{mat90,pol94,and06,chen08}. Mass
transfer is another channel to produce BSs from primordial
binaries. During Roche lobe overflow (RLOF), the less massive star
accretes some material and goes upward along the main sequence, if
it is still a main-sequence star. The accreting component may then
be observed as a BS when it is more massive than the turnoff of
the host cluster \cite{mc64,chen04}. Previous studies show that
both case A and case B mass transfer are only responsible for some
short- and mid-period BSs \cite{leon96,chen04}, which are rare in
some old open clusters, and case C mass transfer may account for
BSs in long-orbital period spectroscopic binaries. During RLOF,
the orbital period will decrease at first, however, if the primary
continues to transfer material to the companion after the primary
is less massive than the secondary, the binary will become wider
and the orbital period will increase. Since dynamical instability
often happens when the mass donor is a giant star, there is little
work to focus on for this channel to BSs.

\begin{figure}
\centerline{\psfig{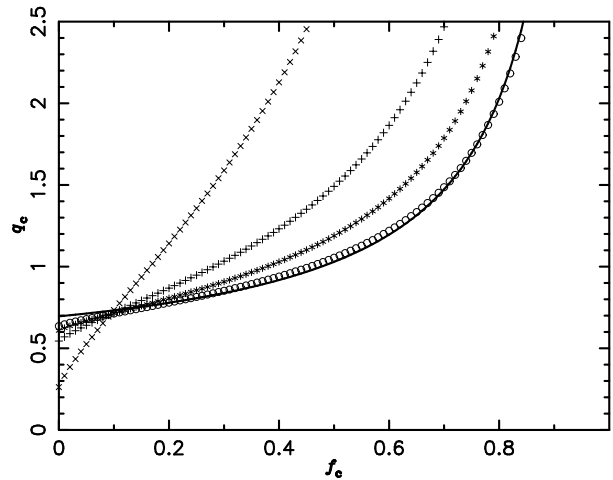}}
\caption{Critical mass ratio $q_{\rm c}$ for dynamically unstable
RLOF. $f_{\rm c}$ is the core mass fraction of the mass donor. The
cross, plus, asterisk and circle lines are for mass transfer
efficiency $\beta =0.25$, 0.5, 0.75 and 1.0, respectively. The
lost mass is assumed to carry away the same specific angular
momentum as pertains to the mass donor. The solid line is from
Webbink (1988) (see equation (1), $f_{\rm c}=M_{\rm c}/M_{\rm 1}$)
for conservative RLOF.} \label{qc}
\end{figure}

As is well known, a fully convective star will increase in radius
with mass loss and decrease in Roche lobe radius if it is more
massive than its accreting companion {\bf \cite{pac65}}. This
means that the mass donor will overfill its Roche lobe by an
ever-increasing amount, leading to mass transfer on a dynamical
time-scale, the formation of a common envelope (CE) and a
spiral-in phase. The critical mass ratio (the mass donor /the
accretor) is about 2/3 from a polytropic model with a polytropic
index $n=1.5$ for conservative mass transfer, indicating that mass
transfer would be dynamically unstable if the mass donor has a
mass larger than 2/3 of the mass of the companion star. If stellar
wind is not included before RLOF to decrease the primary's mass,
case C mass transfer is always dynamically unstable for a binary
with initial mass ratio $q_{\rm i}>1$ \footnote{The primary should
be more massive than the secondary initially from standard stellar
evolution if we expect the binary {\bf to be} composed of a giant
star and a main-sequence companion.} ($q_{\rm i}=M_{\rm 1i}/M_{\rm
2i}$, $M_{\rm 1i}$ and $M_{\rm 2i}$ are the {\bf initial} masses
of the primary and of the secondary, respectively), and {\bf
therefore makes no contribution to BSs.} However, there are some
problems when we use the {\bf above criterion} in a real binary
system for the following facts. (1) Giant stars have large
condensed cores (usually they are degenerate), so they cannot be
modelled by a fully convective star. The critical mass ratio
increases substantially because of the condensed core
\cite{hje87}. They conform to the formula as follows\cite{web88}:

\begin{equation}
q_{\rm c}=0.362+$$1 \over 3(1-M_{\rm c}/M_{\rm 1})$$,
\end{equation}
where $M_{\rm c}$ and $M_{\rm 1}$ are {\bf the core mass and the
total mass, respectively, of the donor as RLOF begins.} (2) The
condition for dynamical instability also depends on the amount of
mass and angular momentum loss during RLOF. Assuming that the lost
mass carries away the same specific angular momentum as pertains
to the mass donor, Soberman et al. \shortcite{sober97} gave a
fitted Roche lobe mass-radius exponent $\zeta_{\rm
L}=-1.7\beta+(2.4\beta -0.25)q$, where $\beta$ is mass transfer
efficiency determined as the mass fraction of the lost mass from
the primary accreted by the secondary. We {\bf may then} obtain
the critical mass ratio $q_{\rm c}$ by setting the adiabatic
mass-radius exponent $\zeta_{\rm S}=\zeta_{\rm L}$, where
$\zeta_{\rm S}$ can be fitted from the data of {\bf numerical
calculations of} Hjellming \& Webbink \shortcite{hje87} and
Soberman et al. \shortcite{sober97} (see also Han et al. 2001).
Fig. \ref{qc} clearly shows the dependence of $q_{\rm c}$ on the
mass transfer efficiency $\beta$. (3) Stellar wind prior to RLOF,
which will be strongly enhanced due to the tidal interaction with
the companion, will increase the core fraction in equation (1) and
allow the system to {\bf stabilize} more easily.

Possibly, a more fundamental problem of such a criterion is that
it does not take into account the detailed dynamics of the mass
transfer process. There are some systems, {\bf observationally,}
which should experience dynamical mass transfer while appearing to
avoid a CE phase \cite{ph92}. Several recent full binary evolution
calculations also show that the simplistic criterion above is not
really appropriate. For example, it is shown by Podsiadlowski et
al. \shortcite{ph02} that mass transfer is dynamically stable for
all giants up to a mass of about $2M_\odot$, in the case of giants
transferring mass to a neutron star of $1.3/1.4M_\odot$. The study
of Han et al. \shortcite{han02} showed that $q_{\rm c}$ may be up
to about 1.3 as $\beta =0$.

In this paper, we will study the critical mass ratio for dynamical
stability in mass transfer from a giant star to a main sequence
companion from detailed binary evolutions, and adopt the results
to estimate the contribution to BSs from the mass transfer. In
section 2, we simply describe the numerical codes we have employed
{\bf in the study. The results} of the critical mass ratio are
shown in section 3. In section 4, we show some examples of
binaries which eventually form long orbital period BSs. The
consequences from Monte Carlo simulations are shown in section 5.
Discussions and conclusions are given in section 6.

\section {Binary Evolution Calculations}
The binary evolution code we employed was originally developed by
Eggleton \shortcite{egg71,egg72,egg73}, which has been updated
with the latest input physics over the last three decades as
described by Han et al.\shortcite{han94} and Pols
\shortcite{pol95,pol98}. Some characteristics of the code, i.e. a
self-adaptive non-Lagrangian mesh, the treatment of both the
convective and semi-convective mixing as diffusion processes, the
simultaneous and implicit solution of both the stellar structure
equations and the chemical composition equations, etc. make the
code very stable and easy to use. The current code uses an
equation of state that includes pressure ionization and Coulomb
interaction, recent opacity tables derived from Rogers \& Iglesias
\shortcite{ir96} and Alexander \& Ferguson \shortcite{alex94} (see
Chen \& Tout, 2007), nuclear reaction rates from Caughlan et
al.\shortcite{cau85} and Caughlan \& Fowler \shortcite{cau88}, and
neutrino loss rates from Itoh et al. \shortcite{it89,it92}.

The ratio of mixing length to the local pressure scaleheight
$\alpha =l/H_{\rm p}$ is set to 2, which fits to the Sun
\cite{pol98}. Convective overshooting is included by incorporating
a condition that mixing occurs in a region with $\nabla _{\rm
r}>\nabla _{\rm a}- \delta _{\rm ov}/(2.5+20\xi +16 \xi ^{2})$,
where $\xi$ is the ratio of radiative pressure to gas pressure and
$\delta_{\rm ov}$ is a specified constant. $\delta_{\rm ov}=0.12$
gives the best fit to the observed systems \cite{sch97}, which
corresponds to an overshooting of about 0.25$H_{\rm p}$. RLOF is
included from the boundary condition
\begin{equation}
$${\rm d}m \over {\rm d}t$$=C{\rm Max}[0,($$r_{\rm star} \over r_{\rm
lobe}$$-1)^3],
\end{equation}
where $r_{\rm star}$ and $r_{\rm lobe}$ are the radii of the mass
donor and its Roche lobe, respectively. ${\rm d}m/{\rm d}t$ gives
the mass loss rate of the primary, and $C$ is a constant. In our
study, we set $C=500M_\odot {\rm yr^{-1}}$, with which RLOF
proceeds steadily and the lobe-filling star overfills its Roche
lobe as necessary but never overfills its lobe by much, typically
$(r_{\rm star}/r_{\rm lobe}-1)<0.001$.

If mass transfer is non-conservative, i.e. $\beta$ is less than
1.0, some matter will {\bf be lost} from the system, taking away
some angular momentum. In our study, the accretor is a
main-sequence star, and it is much {\bf more compact} than a
giant. So we assume that the lost mass carries away the same
specific angular momentum as pertains to that of the accretor,
similar to the case of a compact component. The study of Beer et
al. \shortcite{beer07} shows that this assumption is appropriate
for main-sequence components.

We only follow the evolution of the primary (initially more
massive component) in a binary system when we study the critical
mass ratio, since the structure of the primary is the main factor
for the formation of CE in this phase. {\bf We adopted the solar
metallicity ($Z=0.02$) in our calculations.} As we are mainly
concerned {\bf with BSs which originate from mass transfer,} only
low- and intermediate-mass binaries are studied here. The initial
mass of the primary increases from 1 to $8M_\odot$ by step of
about $\Delta {\rm log}M/M_\odot=0.1$ (i.e. $M_{\rm 1i}$=1.00,
1.26, 1,58, 2.00, 2.51, 3.20, 3.98, 4.95, 6.31 and $7.94M_\odot$,
respectively). When the mass donor evolves to {\bf the} giant
branch, it dramatically expands, and then its radius $R$ may well
represent the evolutionary phases. So we systematically vary the
radius of the mass donor at the onset of RLOF\footnote{\bf The
radius of the primary at the onset of RLOF is actually its Roche
lobe $R_{\rm cr1}$, since the primary is just filling of its Roche
lobe at that time. If the mass ratio $q$ is given, the
corresponding initial orbital separation $A$ can be calculated
from $R_{\rm cr1}/A=0.49q^{2/3}/[0.6q^{2/3}+{\rm ln}(1+q^{1/3})]$
(Eggleton, 1983). For convenience, we only show the radius of the
primary at the onset of RLOF in this paper.} and the mass of the
companion star for each primary mass. If $M_{\rm 1i}<2.00M_\odot$,
the primary has {\bf a} degenerate He core and undergoes {\bf a}
He flash as {\bf the} central He is ignited. Before the He flash,
the star can approach a very high luminosity as well as a very
large radius. So we choose an increase of the radius $\Delta {\rm
log}R/R_\odot$ by {\bf steps} of 0.2 from the base of the first
giant branch (FGB) to the tip of {\bf the} FGB to study the
critical mass ratio. When $M_{\rm 1i} \ge 2.00M_\odot$, He burns
quietly in the center and the radius difference of the primary
from the base to the tip of FGB is not very large. We therefore
set $\Delta {\rm log}R/R_\odot=0.1$ for this case.

\begin{figure}
\centerline{\psfig{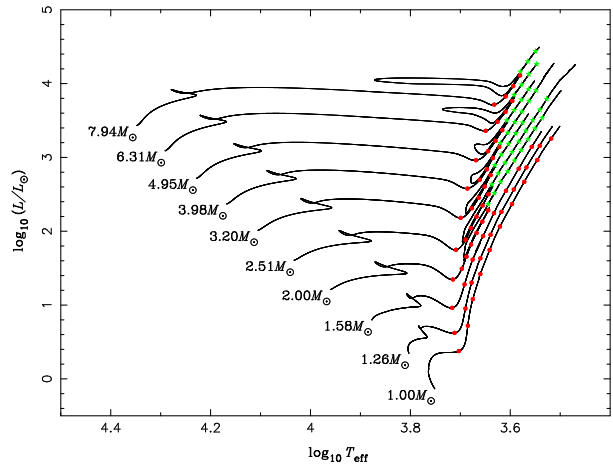}}
\caption{Evolutionary tracks for single stars from 1 to
$8M_\odot$. For primaries in binary systems, the dots (for FGB)
and asterisks (for AGB) indicate their positions at the onset of
RLOF we studied in the paper. Corresponding stellar radii are
presented in Table \ref{1}. The AGB cases for $M_{\rm 1} <2{\rm
M}_\odot$ are not plotted to ensure the distinction of the
figure.} \label{grid}
\end{figure}

If a star has not {\bf filled} its Roche lobe before central He
burning, it will not {\bf fill} its Roche lobe {\bf during} He
burning, because the star will contract after He ignition. We {\bf
therefore} have not considered the He-burning phase in the
studies. After the exhaustion of {\bf its} central He, the star
expands again and its radius may become larger than that at the
tip of {\bf the} FGB. If $M_{\rm 1i}\ge 2M_\odot$, we continue the
study on {\bf the} AGB when the stellar radius equals to that at
the tip of {\bf the FGB}, and increase the radius by {\bf steps}
of $\Delta {\rm log}R/R_\odot=0.1$. Fig. \ref{grid} shows the
positions of the primaries as RLOF begins, corresponding stellar
radii are presented in Table 1.

If $M_{\rm 1i} < 2M_\odot$, the code breaks down when the He flash
occurs. To investigate the case on {\bf the} AGB for {\bf a} mass
donor with $M_{\rm 1} <2.0M_\odot$, we constructed some stellar
models after {\bf the} He flash. Considering that some of the mass
donors on {\bf the} AGB may have evolved from some initially more
massive stars \footnote{For example, a star with $M_{\rm 1i} >
2M_\odot$ becomes less than $2M_\odot$ because of stellar wind,
then {\bf the star has} not passed through a phase with a high
radius before He burning.}, we choose the point at the minimum
stellar radius at the tip of {\bf the} FGB for stars with $M_{\rm
1i} \ge 2M_\odot$ (i.e. ${\rm log}R/R_\odot=1.41$) to start our
studies and increase the radius by {\bf steps} of $\Delta {\rm
log}R/R_\odot=0.1$. The stellar radii and corresponding $q_{\rm
c}$ are listed in Table 2.

\section{The critical mass ratio for a giant mass donor}
\begin{table*}
 \begin{minipage}{170mm}
 \caption{Critical mass ratio for stable RLOF.
$M_{\rm 10}$ is ZAMS mass for the primary; ${\rm log}R$ is the
radius of the primary at the onset of RLOF,
 $q_{\rm c}$ is the critical mass ratio and $\beta$
is the fraction of the mass lost from the primary accreted by the
secondary.}
 \label{tab1}
   \begin{tabular}{ccccccccccccccc}
\hline $M_{\rm 10}$&${\rm log}R$&$q_{\rm c}$&$q_{\rm c}$&$q_{\rm
c}$& $M_{\rm 10}$&${\rm log}R$&$q_{\rm c}$&$q_{\rm c}$&$q_{\rm
c}$&$M_{\rm 10}$&
${\rm log}R$&$q_{\rm c}$&$q_{\rm c}$&$q_{\rm c}$\\
 $\scriptstyle M_\odot$ && $\scriptstyle \beta =0.0$&$\scriptstyle \beta =0.5$&$\scriptstyle \beta =1.0$&
  $\scriptstyle M_\odot$ && $\scriptstyle \beta =0.0$&$\scriptstyle \beta =0.5$&$\scriptstyle \beta
=1.0$&$\scriptstyle M_\odot$ && $\scriptstyle \beta =0.0$&$\scriptstyle \beta =0.5$&$\scriptstyle \beta =1.0$\\
\hline
1.00&0.3048 &1.3943 &1.1927 &1.0607 &2.51&0.9785 &1.4614 &1.2539 &1.0886&2.51&1.6653 &1.2490 &1.0574 &0.8920\\
    &0.5133 &1.2477 &1.0546 &0.8933 &    &1.0867 &1.2757 &1.0842 &0.9279&    &1.7653 &1.2183 &1.0303 &0.8670 \\
    &0.7133 &1.2416 &1.0530 &0.8918 &    &1.1867 &1.2366 &1.0473 &0.8894&    &1.8653 &1.1900 &1.0032 &0.8415\\
    &0.9133 &1.2615 &1.0709 &0.9066 &    &1.2867 &1.2138 &1.0263 &0.8655&    &1.9653 &1.1535 &0.9674 &0.8079\\
    &1.1133 &1.2642 &1.0719 &0.9040 &    &1.3867 &1.1945 &1.0087 &0.8492&    &2.0653 &1.1176 &0.9325 &0.7774 \\
    &1.3133 &1.2545 &1.0595 &0.8902 &    &1.4867 &1.1759 &0.9999 &0.8341&    &2.1653 &1.0847 &0.9002 &0.7471\\
    &1.5133 &1.2228 &1.0312 &0.8660 &&&&&                               &    &2.2653 &1.0639 &0.8776 &0.7259\\
    &1.7133 &1.1901 &0.9968 &0.8324 &3.20&1.2151 &1.5982 &1.3970 &1.2560&    &2.3653 &1.0512 &0.8652 &0.7168\\
    &1.9133 &1.1403 &0.9561 &0.7958 &    &1.3236 &1.3360 &1.1371 &0.9678&    &2.4653 &1.0340 &0.8505 &0.7053 \\
    &2.1133 &1.1131 &0.9257 &0.7696 &    &1.4236 &1.2672 &1.0765 &0.9128&&&&&\\
    &&&&                            &    &1.5236 &1.2200 &1.0338 &0.8738&3.20&1.7076 &1.2668 &1.1075 &0.9235\\
1.26&0.4110 &1.3598 &1.1588 &1.0411 &    &1.6236 &1.1852 &1.0029 &0.8449&    &1.8076 &1.1989 &1.0138 &0.8534\\
    &0.6194 &1.2172 &1.0215 &0.8658 &&&&&                               &    &1.9076 &1.1650 &0.9814 &0.8237 \\
    &0.8194 &1.2226 &1.0272 &0.8685 &3.98&1.4407 &1.7427 &1.5413 &1.3747&    &2.0076 &1.1236 &0.9417 &0.7880\\
    &1.0194 &1.2341 &1.0433 &0.8783 &    &1.5492 &1.3730 &1.1754 &1.0029&    &2.1076 &1.0871 &0.9059 &0.7549\\
    &1.2194 &1.2266 &1.0360 &0.8704 &    &1.6492 &1.2796 &1.0892 &0.9238&&&&&\\
    &1.4194 &1.2089 &1.0174 &0.8532 &    &1.7492 &1.2172 &1.0308 &0.8713&3.98&1.8793 &1.2251 &1.0347 &0.8737\\
    &1.6194 &1.1704 &0.9814 &0.8208 &    &1.8492 &1.1704 &0.9880 &0.8305&    &1.9793 &1.1827 &0.9992 &0.8413  \\
    &1.8194 &1.1289 &0.9399 &0.7835 &&&&&                               &    &2.0793 &1.1399 &0.9581 &0.8043\\
    &2.0194 &1.0871 &0.8983 &0.7471 &4.94&1.6671 &1.7617 &1.5597 &1.3637&    &2.1793 &1.1023 &0.9205 &0.7683\\
    &&&&                            &    &1.7756 &1.3747 &1.1770 &1.0033&    &2.2793 &1.0647 &0.8804 &0.7356\\
1.58&0.5715 &1.3634 &1.1602 &1.0055 &    &1.8756 &1.2719 &1.0823 &0.9161&    &2.3793 &1.0394 &0.8570 &0.7116\\
    &0.7800 &1.2090 &1.0225 &0.8649 &    &1.9756 &1.1952 &1.0116 &0.8512&&&&&\\
    &0.9800 &1.2047 &1.0098 &0.8535 &&&&&                               &4.94&2.0583 &1.2105 &1.0198 &0.8628\\
    &1.1800 &1.2094 &1.0202 &0.8583 &6.31&1.9044 &1.7582 &1.5419 &1.3578&    &2.1583 &1.1619 &0.9792 &0.8249 \\
    &1.3800 &1.1908 &1.0053 &0.8441 &    &2.0129 &1.3547 &1.1576 &0.9844&    &2.2583 &1.1201 &0.9392 &0.7870  \\
    &1.5800 &1.1629 &0.9760 &0.8162 &    &2.1129 &1.2563 &1.0689 &0.9034&    &2.3583 &1.0823 &0.9013 &0.7524\\
    &1.7800 &1.1174 &0.9310 &0.7764 &    &2.2129 &1.1548 &0.9740 &0.8178&&&&&\\
    &1.9800 &1.0771 &0.8886 &0.7393 &&&&&                               &6.31&2.2613 &1.1864 &0.9971 &0.8392\\
    &&&&                            &7.94&2.1091 &1.6967 &1.4669 &1.2748&    &2.3613 &1.1438 &0.9622 &0.8090 \\
1.91&0.7239 &1.3905 &1.1847 &1.0165 &    &2.2176 &1.2971 &1.1004 &0.9288&    &2.4613 &1.0985 &0.9177 &0.7686 \\
    &0.9324 &1.2140 &1.0269 &0.8695 &    &2.3176 &1.2306 &1.0457 &0.8804&    &2.5613 &1.0651 &0.8841 &0.7357\\
    &1.1324 &1.1971 &1.0078 &0.8504 &    &2.4176 &1.1247 &0.9476 &0.7924&&&&&\\
    &1.3324 &1.1816 &0.9977 &0.8369 &&&&&                               &7.94&2.4438 &1.1624 &0.9764 &0.8201\\
    &1.5324 &1.1652 &0.9803 &0.8217 &&For& AGB&case&                    &    &2.5438 &1.1235 &0.9418 &0.7913 \\
    &1.7324 &1.1223 &0.9387 &0.7838 &2.00&1.4175 &1.2764 &1.0819 &0.9130&    &2.6438 &1.1019 &0.9203 &0.7700\\
    &&&&                            &    &1.5175 &1.2661 &1.0730 &0.9063 \\
2.00&0.7672 &1.3710 &1.1779 &1.0065 &    &1.6175 &1.2388 &1.0474 &0.8805 \\
    &0.8759 &1.2447 &1.0530 &0.9035 &    &1.7175 &1.2068 &1.0171 &0.8526 \\
    &0.9757 &1.2261 &1.0298 &0.8745 &    &1.8175 &1.1698 &0.9816 &0.8210\\
    &1.0757 &1.2039 &1.0157 &0.8614 &    &1.9175 &1.1329 &0.9470 &0.7890\\
    &1.1757 &1.1980 &1.0093 &0.8521 &    &2.0175 &1.0971 &0.9090 &0.7546\\
    &1.2757 &1.1890 &1.0030 &0.8446 &    &2.1175 &1.0901 &0.9045 &0.7462\\
    &1.3757 &1.1786 &0.9936 &0.8355 &    &2.2175 &1.0734 &0.8844 &0.7313\\
\hline \label{1}
\end{tabular}
\end{minipage}
\end{table*}

\begin{table*}
 \begin{minipage}{170mm}
 \caption{Critical mass ratio for stable RLOF for AGB mass donors with mass less than $2M_\odot$.
$M_{\rm 1}$ is the mass of the primary, ${\rm log}R$ is the radius
of the primary at the onset of RLOF,
 $q_{\rm c}$ is the critical mass ratio and $\beta$
is the fraction of the mass lost from the primary accreted by the
secondary. }
 \label{tab1}
   \begin{tabular}{ccccccccccccccc}
\hline
${\rm log}R$&$M_{\rm 1}=1.00M_\odot$&&&&$M_{\rm 1}=1.26M_\odot$&&&&$M_{\rm 1}=1.60M_\odot$&&&\\
\hline $R_\odot$   &$\beta=0.$ &$\beta=0.5$&$\beta=1.$&&$\beta=0.$
&$\beta=0.5$&$\beta=1.$&&$\beta=0.$ &$\beta=0.5$&$\beta=1.$&\\
\hline
1.4096& 2.0406& 1.7919& 1.5816&& 1.7504 &1.4589 &1.2640&& 1.5018& 1.3020& 1.1192\\
1.5096& 1.9040& 1.6705& 1.4550&& 1.5912 &1.3709 &1.2294&& 1.4360& 1.2290& 1.0533\\
1.6096& 1.8061& 1.5761& 1.3728&& 1.5302 &1.3154 &1.1481&& 1.3954& 1.1904& 1.0157\\
1.7096& 1.7154& 1.4891& 1.3096&& 1.4791 &1.2676 &1.0833&& 1.3617& 1.1608& 0.9834 \\
1.8096& 1.6223& 1.4045& 1.2144&& 1.4142 &1.2078 &1.0256&& 1.3059& 1.1100& 0.9362\\
1.9096& 1.5773& 1.3260& 1.1437&& 1.3544 &1.1529 &0.9759&& 1.2575& 1.0629& 0.8940\\
2.0096& 1.4503& 1.2448& 1.0076&& 1.2937 &1.0970 &0.9228&& 1.2058& 1.0146& 0.8486\\
2.1096& 1.3590& 1.1681& 0.9986&& 1.2307 &1.0382 &0.8698&& 1.1540& 0.9631& 0.8053\\
2.2096& 1.2955& 1.1014& 0.9327&& 1.1827 &0.9904 &0.8290&& 1.1167& 0.9290& 0.7730\\
2.3096& 1.2553& 1.0649& 0.8849&& 1.1562 &0.9658 &0.8057&& 1.0976& 0.9094& 0.7542\\
\hline \label{q2}
\end{tabular}
\end{minipage}
\end{table*}

\begin{figure}
\centerline{\psfig{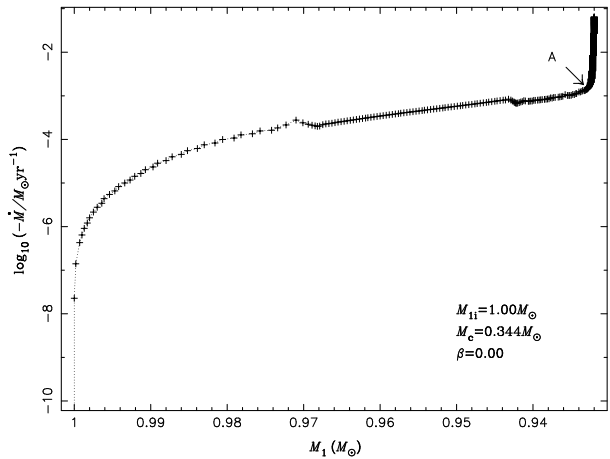}}
\caption{Mass transfer rate vs the primary mass for a binary of
$M_{\rm 1i}=1.00 M_\odot$. Mass transfer efficiency $\beta =0.0$
and the core mass is $0.344M_\odot$. Point A gives the place at
which mass transfer dramatically increases because of dynamical
instability.} \label{du}
\end{figure}

As mentioned in section 1, dynamically unstable RLOF often occurs
when the mass donor is a giant at the onset of mass transfer. Fig.
\ref{du} shows the mass transfer rate vs the mass of the primary
for a typical binary undergoing dynamical unstable mass transfer.
The primary is $1.00M_{\odot}$ with a core mass $M_{\rm
c}=0.344M_{\odot}$ at the onset of RLOF, and $\beta =0$ during the
RLOF. From the figure we see that mass transfer initially occurs
on a thermal time-scale. The mass transfer rate, $\dot M$, rises
quickly to a level of $10^{-5}M_\odot {\rm yr^{-1}}$ at first and
continues to grow more slowly {\bf later}. After {\bf the primary
has lost about $0.04M_\odot$} (point A), the system encounters
dynamical instability. The behaviour here is similar to that of
radiative mass donors with a mass ratio greater than 3
\cite{han06}, but it cannot be explained by the evolution of the
entropy profile as in the case of {\bf a} mass donor with a
radiative envelope. It is related to the evolution of binary
parameters, which strongly depend on the angular momentum loss
associated with the mass loss from the system. Furthermore, in
comparison to the delayed dynamical instability in the radiative
mass donor, the primary with a convective envelope here loses much
less mass during the delayed time and probably has little {\bf
influence} on the final {\bf outcome}.

The results of our calculations are summarized in Tables \ref{1}
and \ref{q2}. For each ZAMS, we present the radius of the primary
${\rm log}R/R_\odot$ at the onset of RLOF and the critical mass
ratio $q_{\rm c}$ for dynamically stable RLOF from different mass
transfer efficiency $\beta$. The results demonstrate that RLOF is
{\bf probably} dynamically stable even if the mass donor is more
massive than the secondary for $\beta < 1$. In particular, when
the mass donor is less than $2M_\odot$ and fills its Roche lobe on
{\bf the} AGB, RLOF may be dynamically stable even for $\beta =
1$. Since the core is not very degenerate and the envelope is not
fully convective when the star is at or near the base of {\bf the}
FGB, $q_{\rm c}$ is much larger at the base of {\bf the} FGB than
in the following evolutionary phases for each $\beta$.

\begin{figure}
\centerline{\psfig{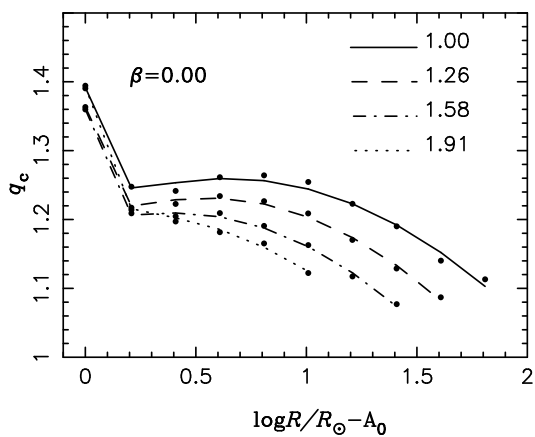}}
\caption{Critical mass ratio for low-mass binaries of FGB+MS, i.e.
$M_{\rm 1i}=1.00$, 1.26, 1.60 and $1.90M_\odot$. The lines are
from the fitting formulae of equations (4) to (6).} \label{low-qc}
\end{figure}

\begin{figure*}
\centerline{\psfig{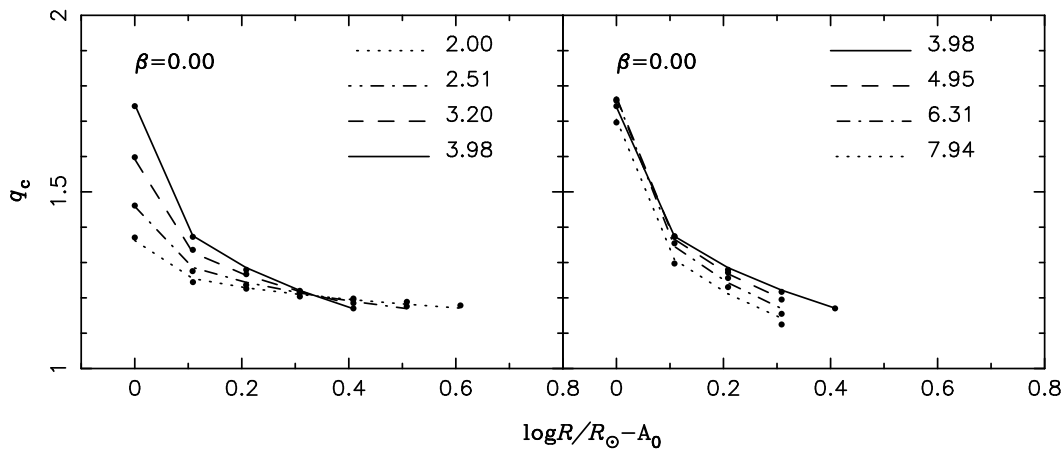}}
\caption{Similar to figure \ref{low-qc}, but for
intermediate-massive binaries, i.e. $M_{\rm 1i}=2.00$, 2.50, 3.20,
4.00, 5.00, 6.30 and $8.00M_\odot$. The lines are from equations
(5) and (6).} \label{mid-qc}
\end{figure*}

\begin{figure}
\centerline{\psfig{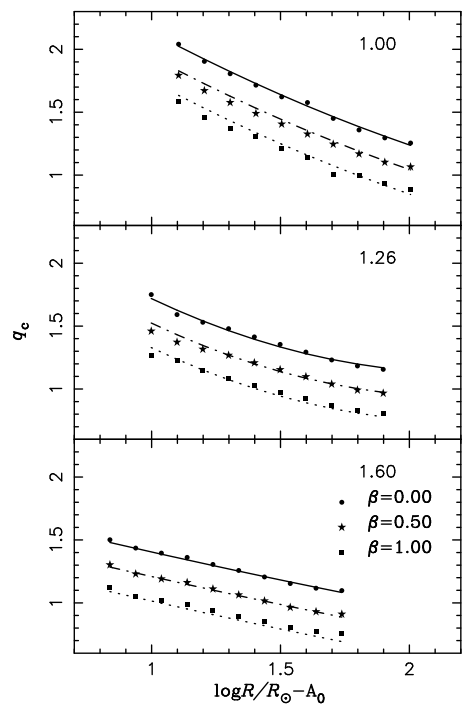}}
\caption{Critical mass ratio for low-mass binaries when the
primary is on the AGB. The lines are from the fitting formulae of
equations (6) and (7).} \label{agb2}
\end{figure}

\begin{figure}
\centerline{\psfig{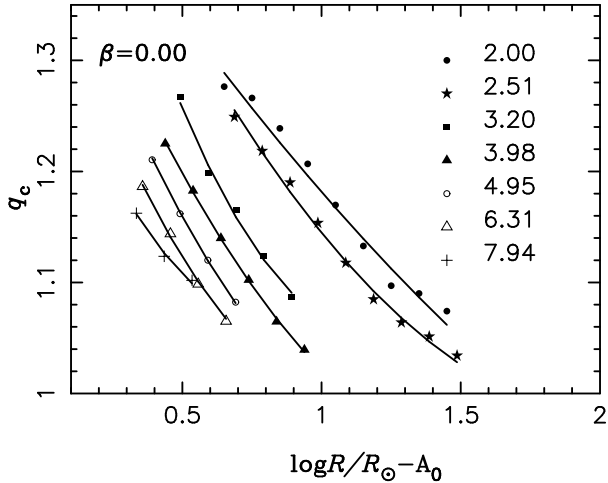}}
\caption{Critical mass ratio for intermediate-mass binaries when
the primary is on the AGB. The lines are from the fitting formulae
of equations (6) and (7).} \label{agb}
\end{figure}

Figs. \ref{low-qc} and \ref{mid-qc} present $q_{\rm c}$ vs ${\rm
log}R/R_\odot-A_{\rm 0}$ {\bf for low-mass and intermediate-mass
binaries,} respectively, when the mass donors are on the FGB. Here
$A_{\rm 0}$ is {\bf the log of the radius of the primary at the
base of the FGB}. We include the result of $M_{\rm 1i}=1.9M_\odot$
in Fig. \ref{low-qc} (see also Table \ref{1}). The results for the
mass donor on {\bf the} AGB are showed in Figs. \ref{agb2} and
\ref{agb}. In all of the figures, we see that a more evolved star
(with a larger stellar radius at the onset of RLOF) is less stable
for RLOF, especially when the mass donor is on {\bf the} AGB at
the onset of mass transfer. The reasons are as follows. For more
evolved binaries, the evolutionary time-scale is shorter and hence
the mass transfer rate higher than those for the less-evolved
ones. However $q_{\rm c}$ is non-monotonic initially in
Fig.\ref{low-qc} because the core is not very degenerate and the
envelope is not yet full convective in this phase.

The cases at the last point on {\bf the} FGB and at the first
point on {\bf the} AGB give a different tendency as described
above, since the two points are on different giant branches, and
{\bf the stars have different cores, i.e. a non-degenerate He core
on the FGB while a possibly degenerate CO core on the AGB.}
Meanwhile, the degree of convection is also different at the two
points.

As shown in Fig.\ref{mid-qc}, there is a threshold for $q_{\rm c}$
when the initial primary mass $M_{\rm 1i}=3.98M_\odot$. When
$M_{\rm 1i}\le3.98M_\odot$, $q_{\rm c}$ systematically increases
with the initial primary's mass while it is opposite when $M_{\rm
1i}\ge3.98M_\odot$. This phenomena is likely revelant to the
degenerate degree of the core. We examined the degeneracy
parameter $\psi$ in the calculations and found that $\psi$ becomes
negative just at $M_{\rm 1i}=3.98M_\odot$.

\begin{figure}
\centerline{\psfig{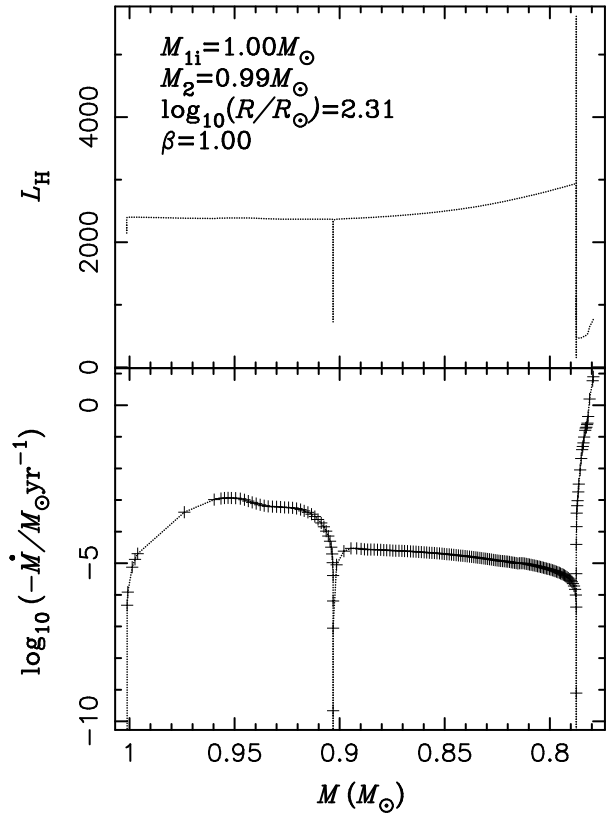}}
\caption{Mass transfer rate as well as H-burning luminosity vs the
primary mass for a binary when the mass donor is on the AGB. }
\label{agb3}
\end{figure}

We know that, when central He is exhausted in a star, there {\bf
exists} a He-burning shell and a H-burning shell. The H-burning
shell extinguishes first when it burns outwards, and then the
He-burning shell also extinguishes. The star begins to contract
and the temperature increases until H is ignited. With the
increase of temperature, He is also ignited but in a very thin
shell. The He ignition in a very thin shell is unstable, {\bf and}
leads to a {\bf dramatic} expanding of the star. After the
expansion, He-burning becomes stable and the star then has two
stable burning shells again, and so on. This behaviour is known as
{\bf the} thermal pulse. The thermal pulse may result in a sharp
increase of mass transfer rate and leads dynamical instability
{\bf to be more likely to occur} such as in  Fig. \ref{agb3}. In
principle, if the mass donor has two burning shells, the thermal
pulse will likely affect the final results. For the cases we
studied, however, the influences seem {\bf small} and can be
neglected.

\begin{figure}
\centerline{\psfig{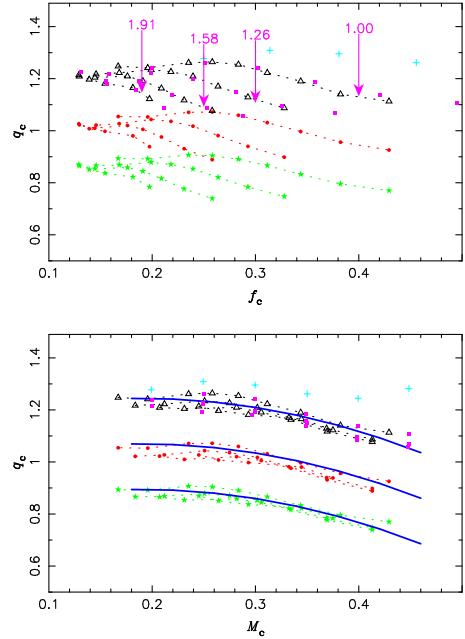}}
\caption{The dependencies of critical mass ratio $q_{\rm c}$ on
the core mass fraction $f_{\rm c}$ and on the core mass $M_{\rm
c}$ for low-mass binaries, i.e the initial primary mass $M_{\rm
1i}=1.00$, 1.26, 1.6 and 1.9$M_\odot$, respectively. The
triangles, dots and asterisks are for mass transfer efficiency
$\beta =0.$, 0.5 and 1.0, respectively. The squares and pluses are
from Han et al (2002), where stellar wind is included before RLOF
and $\beta=0.$ The pluses are for binaries with a primary mass
equal to $0.8M_\odot$, and the squares for $M_{\rm 1i} \ge
1.00M_\odot$. The solid lines in the bottom panel are from
equations (8) and (9).} \label{low-fit}
\end{figure}

To simply use this results, we fitted $q_{\rm c}$ from our
calculations. The influence from $\beta$ is nearly linear and can
be fitted as

\begin{equation}
q_{\rm c}=q_{\rm c0}-c_0\beta,
\end {equation}

For {\bf the} FGB,
\begin{equation}
q_{\rm c0}=c_1+c_2({\rm log}R/R_\odot-A0)+c_3({\rm
log}R/R_\odot-A0)^2,
\end {equation}
when $M_{\rm 1i}<2.00M_\odot$ and ${\rm log}R/R_\odot-A0 > 0.2$.
In other cases,
\begin{equation}
 q_{\rm c0}=c_4+c_5({\rm log}R/R_\odot-A0)^{1/3}.
\end {equation}
Here
\begin{equation}
c_{\rm i}=c_{\rm i,1}+c_{\rm i,2}M_{\rm 1}+c_{\rm i,3}M_{\rm 1}^2.
\end {equation}
The values of $c_{i,j}$ and $c_0$ are listed in Table \ref{c-fgb}
for different cases.

For the case of {\bf the} AGB,
\begin{equation}
q_{\rm c0}=c_1+c_2({\rm log}R/R_\odot-A0)+c_3({\rm
log}R/R_\odot-A0)^2.
\end {equation}

The {\bf coefficients} $c_i$ and $c_0$ are listed in Table
\ref{c-agb}. One may send {\bf a} request to {\it
xuefeichen717@hotmail.com} {\bf for the FORTRAN code of all the
formulae used in this paper.}

\begin{table}
 \begin{minipage}{80mm}
 \caption{The coefficients $c_{i,j}$ and $c_0$ for various cases when the mass donor is on the FGB (from equations(7) to (10)).
  Here $r={\rm log}R/R_\odot-A0$.}
 \label{tab1}
   \begin{tabular}{cccccc}
\hline
cases &$i$&$j=1$&$j=2$&$j=3$&$c_0$\\
\hline
$M_{\rm 1} <2.00M_\odot$   &$1$&1.4418&-0.3477&0.1180&0.37\\
 \&                        &$2$&-0.08252&0.4180&-0.1873&\\
  $r \ge 0.2$              &$3$&0.08967&-0.3129&0.1079\\
\hline
$M_{\rm 1} <2.00M_\odot$   &$4$&1.7141&-0.4919&0.1695&0.37\\
\& $r<0.2$                     &$5$&-0.3978&0.2620&-0.1101&\\
\hline
$M_{\rm 1}\ge2.00M_\odot$ &$4$&0.9989&0.1755&0.003139&0.35\\
\& $M_{\rm 1} <3.98M_\odot$   &$5$&0.2740&-0.2353&-0.007348&\\
\hline
$M_{\rm 1} \ge 3.98M_\odot$&$4$&1.3840&0.1389&-0.0124&0.37\\
                            &$5$&0.05596&-0.2602&0.02045\\
\hline
\label{c-fgb}
\end{tabular}
\end{minipage}
\end{table}

\begin{table}
 \begin{minipage}{80mm}
 \caption{The coefficients $c_{i,j}$ and $c_0$ for various cases when the mass donor is on the AGB. }
 \label{tab1}
   \begin{tabular}{cccccc}
\hline
cases &$i$&$j=1$&$j=2$&$j=3$&$c_0$\\
\hline
$M_{\rm 1} <2.51M_\odot$   &$1$&1.0462 & 0.3115  &-0.03702&0.34\\
                            &$2$&-0.6179 &0.5193  &-0.2027\\
                            &$3$&0.7346 &-0.7568 &0.2074\\
\hline
$M_{\rm 1}\ge2.51M_\odot$ &$1$&-0.5270 & 1.4289 &-0.2325&0.36 \\
      \&                  &$2$&7.1432 &-4.9954  &0.7625\\
$M_{\rm 1} <3.98M_\odot$  &$3$&-4.9258 &3.2653 &-0.4966\\
\hline
$M_{\rm 1} \ge 3.98M_\odot$&$1$&1.4681  &0.01922 &-0.004501&0.36 \\
                            &$2$&0.1092  &-0.2844 &0.02368\\
                            &$3$&-0.1971 &0.1226  &-0.006051\\
\hline
$M_1<2.00M_\odot$           &$1$&0.02442 &7.0840  &-3.6961& 0.39\\
                            &$2$&9.6761  &-19.0892 &7.9462&  \\
                            &$3$&-4.5199 & 7.8189  &-3.1087\\
 \hline \label{c-agb}
\end{tabular}
\end{minipage}
\end{table}

To compare with the results of the polytropic model, we present
the dependence of $q_{\rm c}$ on the core mass $M_{\rm c}$ as well
as on the core mass fraction $f_{\rm c}$ for low-mass binaries in
Fig. \ref{low-fit}, ignoring the points at the base of {\bf the
FGB} for each mass. Here the core mass is defined as the mass
within the hydrogen mass fraction less than 0.1 \footnote{ It is a
little bit difficult to determine the core in a real star as
described in the polytropic model. However, the core mass defined
here {\bf is a close approximation to that defined in the
polytropic model.}}. From the figure, we see a clear relation
between $q_{\rm c}$ and $M_{\rm c}$, but not between $q_{\rm c}$
and $f_{\rm c}$. $q_{\rm c}$ may be fitted as follows:
\begin{equation}
q_{\rm c}=q_{\rm c0}-0.35\beta
\end {equation}
where
\begin{equation}
q_{\rm c0}=1.142+1.081M_{\rm c}-2.852M_{\rm c}^2.
\end {equation}

The studies above have not included the influences from {\bf the}
stellar wind.  Han et al. \shortcite{han02} included the stellar
wind prior to RLOF when they studied the $q_{\rm c}$ for low-mass
binaries on the FGB and $\beta=0$. We have included their results
in Fig. \ref{low-fit}. The pluses are for binaries with $M_{\rm
1i}=0.8M_\odot$\footnote{We have not studied the case for $M_{\rm
1i}=0.8M_\odot$ in this paper, because the time-scale is too long
for a star with this mass to evolve from zero-age main sequence to
giant branch. {\bf Meanwhile, it might be more likely that
low-mass binaries (i.e the initial primary less than $1M_\odot$)
contribute to blue stragglers via coalescence induced by angular
momentum loss \cite{chen08}, while not from the mass transfer
between a giant and a MS.}} and the squares for $M_{\rm 1i}\ge
1.00M_\odot$ (see Table 3 in their paper). We see that all the
squares are well along the fitting line of $\beta=0$, indicating
that stellar wind has little influence on $q_{\rm c}$. However,
the mass ratio will decrease due to stellar wind, and RLOF will be
more stable, e.g. the mass ratio possibly becomes less than
$q_{\rm c}$  at the onset of RLOF and the binary will undergo a
stable mass transfer. For the case of $M_{\rm 1i}=0.8M_\odot$,
$q_{\rm c}$ from Han et al. \shortcite{han02} is obviously larger
than the fitting value, which means that RLOF is more stable in
binaries with this {\bf primary} mass. Since the binaries with
$M_{\rm 1i}=0.8M_\odot$ have a larger $f_{\rm c}$ in comparison to
those with $M_{\rm 1i} \ge 1.0M_\odot$ \footnote{{In the study of
Han et al. \shortcite{han02}, the mass donor has a similar core
mass but a different stellar mass at the onset of RLOF. Therefore,
a high stellar mass means a smaller core mass fraction $f_{\rm
c}$. See Table 3 in their paper for details.}}, the high $q_{\rm
c}$ for $M_{\rm 1i}=0.8M_\odot$ here likely indicates some
contributions from $f_{\rm c}$, as is possibly concealed by that
from $M_{\rm c}$ for binaries with $M_{\rm 1i} \ge 1.00M_\odot$
(see the discussion in section 6).

\section{examples of three binaries resulting in BSs}
In this section, we present the detailed evolutions for three
binaries, which are selected mainly {\bf for the reason that} we
are interested in BSs in the old open cluster M67, where many BSs,
including several long-period BSs, have been observed.  {\bf M67
has a metallicity similar to that of the Sun \cite{hob91,fri93}.
There are some researches on the age of M67. It may range from
3.2{\underline +}0.4 Gyr \cite{bb03} to 6.0 Gyr \cite{jp94}. The
study of VandenBerg \& Stetson \shortcite{vs04} derived an age of
4.0 Gyr. In the N-body model of this cluster \cite{hur05}, the
authors investigated the behaviour around 4 Gyr. So we also
consider that its age is about 4 Gyr, indicating that the mass of
the turnoff $M_{\rm to}$ is about $1.2$ to $1.3M_\odot$, which is
the main factor of the cluster we will consider as we choose the
binary samples.} Basic requirements for the binaries which may
contribute to BSs in a cluster via a giant transferring mass to a
MS companion are for the masses of both components, i.e. the
primary should be more massive than the turnoff to ensure that it
has left the main sequence and the secondary should be less than
the turnoff if it is still on the main sequence at the cluster
age. The fact that the secondary should accrete certain material
before it becomes a BS will give a further constraint on the
binary parameters. Obviously, $\beta =0$ is ruled out since no
material is accreted by the secondary in this case. In the
following of this section, we will give three examples and present
the evolutionary results in details.

Example 1: RLOF is dynamically stable for initial parameters. As
shown in section 3, $q_{\rm c}$ decreases with the mass transfer
efficiency $\beta$. This means that, in order to ensure that
$\beta$ is large enough to increase the secondary's mass to be
larger than the turn-off, the mass ratio $q$ should be as small as
possible. However, $q$ should be larger than unity to confirm that
mass transfer occurs between a giant star and a MS companion. We
therefore choose a binary of $1.3 +1.2M_\odot$ ($q=1.08$).
Meanwhile, since the orbital period increases with the core mass
(or the radius) of the mass donor at the onset of RLOF, a large
core mass (or stellar radius) is necessary for long orbital period
BSs. However, when $M_{\rm c}>0.4M_\odot$, $q_{\rm c}$ is less
than 1.0 even for $\beta=0$. We therefore set $M_{\rm
c}=0.35M_\odot$ (${\rm log}R/R_\odot=1.7503$) at the onset of
RLOF. From equations (8) and (9), $\beta =0.16$ for $q_{\rm
c}=1.08$. We then set $\beta=0.1$ in the calculation.

Example 2: RLOF is dynamically unstable for initial parameters,
but it will be stable after some mass loss of the primary by
stellar wind prior to RLOF. Stellar wind is included only {\bf on}
the giant branch by the mode of Reimers' \shortcite{rei75}:
\begin{equation}
\dot M_{\rm wind}=4\times10^{-13} \eta RL/M,
\end{equation}
where $\eta $ is a {\bf dimensionless} factor. The same binary as
example 1 was examined here but with $\beta=0.5$. {\bf In general,
$\eta $ is 1/4 \cite{ren81,ir83,car96}. However it is expected by
many authors that giant stars have much higher mass loss rates
than that of Reimers, especially when the stars are far away from
the base of giant branch (e.g. Bloecker 1995). Furthermore, tidal
interaction between the two components of the binary also
increases the mass loss rate. So we set $\eta =2.5$ \footnote{We
also examined the cases for $\eta = 1/4$ and 1, and found that
both of them are too small to strip enough mass away to lead to
$q<q_{\rm c}$ as RLOF begins.} in the calculation.} When RLOF
begins ($M_{\rm c}=0.355M_\odot$), $M_{\rm
1}=1.11M_\odot$($q=M_{\rm 1}/M_{\rm 2}=0.93$). We switch the
stellar wind off once the mass transfer rate exceeds the value
given by eq(10) and have not included it after RLOF. From eqs(8)
and (9), we get $q_{\rm c}=0.99$ for $\beta=0.5$ and $M_{\rm
c}=0.35M_\odot$. So RLOF is dynamically stable as it begins.

Example 3: RLOF is dynamically unstable for initial parameters,
but it becomes stable after the initially dynamically unstable
mass transfer, if we assume that CE has not developed during this
phase according to the study of Beer et al. \shortcite{beer07}.
Beer et al. \shortcite{beer07} argued that a wide range of systems
avoid CE phase because mass transfer is super-Eddington even for
non-compact companions. The accretion energy released in the rapid
mass-transfer phase strips away a large fraction of the giant's
envelope, reducing the tendency to dynamical instability and
merging. More constraints are necessary to determine {\bf whether}
the CE is formed or not in this case, but we have little knowledge
{\bf about} it. For simplicity, we assume that CE has not
developed in the binary we study here, i.e. in a binary with
$1.6+1.1M_\odot$. The code cannot work for the initial dynamically
unstable mass transfer because of the high mass transfer rate
$\dot M$. We therefore artificially limit the highest $\dot M$ to
be $1\times 10^{-4} M_\odot yr^{-1}$, and the accretion rate of
the secondary $\dot M _{\rm a}$ is equal to $\dot M$ as $\dot M <
1\times 10^{-5} M_\odot {\rm yr}^{-1}$, and equal to $1 \times
10^{-5}M_\odot {\rm yr}^{-1}$ as $\dot M \ge 1\times 10^{-5}
M_\odot {\rm yr}^{-1}$. From this assumption, $\beta$ ranges from
1.0 to 0.1, then to 1.0 again during the whole RLOF {\bf period}.
The maximum accretion rate of the MS companion here, i.e. $1
\times 10^{-5}M_\odot {\rm yr}^{-1}$, is comparable with the
typical value of symbiotic stars \cite{scott84}.

\begin{figure}
\centerline{\psfig{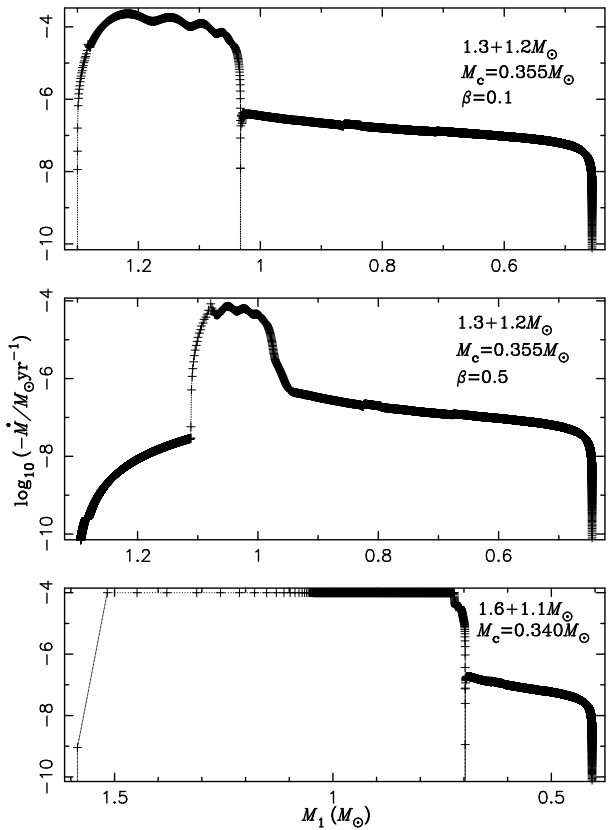}}
\caption{Mass loss rate vs the mass of the primary for the three
binaries in the text. The system parameters are presented in each
panel. Stellar wind is not included in panels (a) and (c), while
it is included at the giant branch in panel (b) with the mode of
Reimer ($\eta = 2.5$). During the initially dynamically unstable
phase in panel (c), we set the highest mass loss rate of the
primary to be $1 \times 10^{-4}M_\odot {\rm yr}^{-1}$ (see the
text for details).} \label{dmt3}
\end{figure}

Figure \ref{dmt3} presents the mass loss rate to the total mass of
the primary for the three binaries. In these figures we see that,
after the initial rapid mass loss, the primary loses its material
at a rate {\bf of} about $10^{-6}$ to $10^{-8} M_\odot {\rm
yr}^{-1}$. Generally this slow mass transfer occurs after {\bf the
inversion of the mass ratio}, i.e. the primary is less massive
than the secondary. During the slow mass transfer phase, the
radius of the primary {\bf again increases} but with a much lower
level in comparison to that in the rapid mass loss phase. At the
same time, the Roche lobe radius of the primary increases with
mass loss, since the orbital period becomes longer and longer
after the {\bf inversion} of $q$. So if CE has not developed in
the dynamically unstable phase in a binary such as example 3, it
is very likely that CE will never develop in the binary. The
material lost from the primary in this phase will {\bf be lost}
from {\bf the} system in a way similar to stellar wind, and RLOF
eventually terminates after most of the primary's envelope is
lost. Note that the mass loss of the system here is different from
the ejection of CE. The former needs no orbital energy but the
latter needs some. So binaries avoiding CE have long orbital
periods.

\begin{figure}
\centerline{\psfig{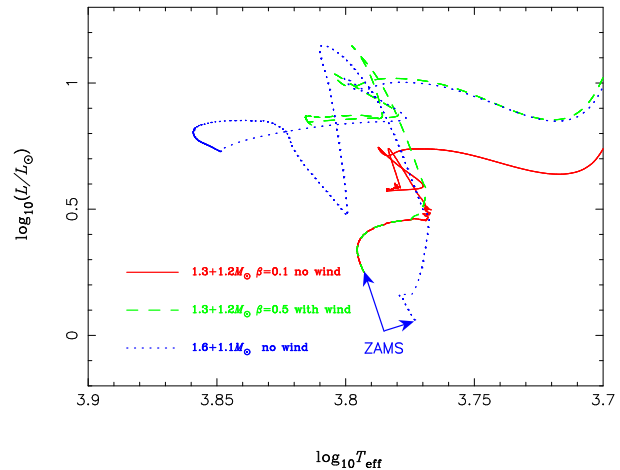}}
\caption{The evolutionary tracks for the secondaries of the three
binaries we examined in the text.} \label{s2}
\end{figure}

The behaviours of the secondaries of the three binaries are
presented in Fig. \ref{s2}. Since the secondaries cannot {\bf
accrete} the mass from the primary in a very short time, they have
left thermal equilibrium and evolve in a way similar to pre-MS
stars during the initial phase instead of going upward along the
main sequence. Due to {\bf the} long orbital periods, the
secondaries {\bf do not overfill} their Roche lobes. They finally
become normal main-sequence stars with a higher He fraction in the
envelope when new thermal equilibrium is established, and evolve
similarly to normal single stars with {\bf those} masses. These
rejuvenated stars have much longer timescales (including the phase
of the secondaries with lower masses before accretion) on the main
sequence and have the possibility to be BSs.

The main characteristics of the three binaries are listed in Table
\ref{2}, from which we may obtain some clues on their parent
binaries and evolutionary histories. For example, although the
secondary in example 2 has a mass similar to that in example 3
after accretion, the latter is much bluer (Fig. \ref{s2}) when
both of them return to thermal equilibrium since the latter is
much less evolved at the onset of RLOF. The two secondaries have
different orbital periods, different compact components and
different lifetimes etc., and appear in different age of the
cluster, as shown in Table \ref{2}.

\begin{figure}
\centerline{\psfig{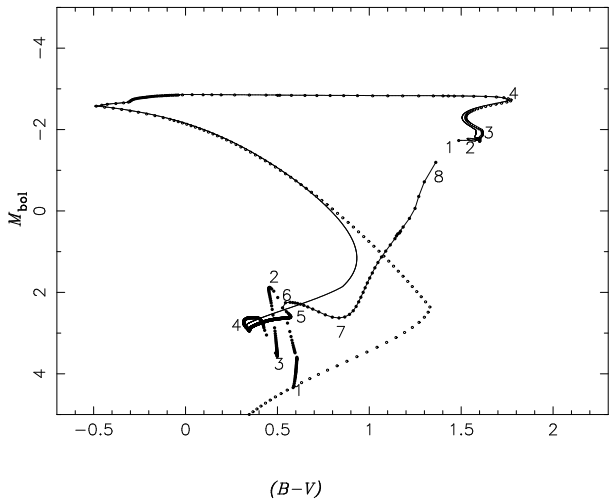}}
\caption{The colour and magnitude evolutions for the individual
components of the binary $1.6+1.1M_\odot$ as well as for the whole
system. The early evolutions of both components prior to RLOF are
not plotted in the figure. The circle and dotted lines are for the
primary and the secondary, respectively, and the solid line is the
combination of the two components. {\bf The numbers are the time
sequences of the evolutions.}} \label{color}
\end{figure}

As an example, we give a colour-magnitude diagram of both of the
components as well as the binary system for example 3 in
Fig.\ref{color}. The combined colour and magnitude of the binary
result from add-and-subtract calculations of the luminosity of
both components (see Appendix A). The evolutions of the components
prior to RLOF have no differences from a standard stellar
evolution and are not plotted in the figure. During the whole mass
transfer phase the system resembles the characteristics of the
primary because the primary is much more luminous than the
secondary, but the {\bf timescale} is very short. Once the primary
becomes {\bf cooler}, i. e. less luminous than its component, the
binary is hardly {\bf distinguishable} from the main-sequence
component. It is very difficult to find the clues from the compact
component, even from {\bf spectral} observations because of the
long orbital period.

\begin{table*}
 \begin{minipage}{150mm}
 \caption{The characteristics for three different evolutionary examples in
the text.$\eta$ = the factor of Reimer's wind, $M_{\rm c}$ = the
core mass at the onset of RLOF, $\beta$ = the mass fraction of the
lost matter of the primary accreted by the secondary, $t_{\rm
RLOF}$ = the age at the onset of RLOF, $t_{\rm MS}$ = the age of
the secondary when it terminates its main sequence $P_{\rm e}$,
$M_{\rm 1e}$, $M_{\rm 2e}$ are the orbital period, the mass of the
primary and the mass of the secondary at the end of RLOF. }
 \label{tab2}
   \begin{tabular}{cccccccccc}
\hline
&binary &stellar wind &$M_{\rm c}$&$\beta$&$P_{\rm e}$&
$t_{\rm RLOF}$&$t_{\rm MS}$& $M_{\rm 1e}$ & $M_{\rm 2e}$\\
 &$\scriptstyle M_\odot$ & $\eta$&$\scriptstyle M_\odot$ & &
   $\scriptstyle {\rm days}$&  $\scriptstyle {\rm Gyrs}$&
$\scriptstyle {\rm Gyrs}$& $\scriptstyle M_\odot$&
 $\scriptstyle M_\odot$\\
\hline
example 1 & 1.3+1.2 & 0.0 & 0.356 & 0.1 & 803 & 4.71 & 5.45 & 0.46 & 1.28\\
example 2 & 1.3+1.2 & 2.5 & 0.356 & 0.5 & 722 & 4.71 & 5.45 & 0.44 & 1.53\\
example 3 & 1.6+1.1 & 0.0 & 0.340 & -   & 439 & 2.47 & 4.57 & 0.41 & 1.50\\
\hline
\label{2}
\end{tabular}
\end{minipage}
\end{table*}

\section{Monte Carlo Simulation}
To investigate BSs from mass transfer between giants and MS
companions, we performed a Monte Carlo simulation for a sample of
$10^6$ binaries (very wide binaries are actually single stars). A
single starburst is assumed in the simulation, i.e. all the stars
have the same age and metallicity ($Z=0.02$). The initial mass
function (IMF) of the primary, the initial mass ratio distribution
and the distribution of initial orbital separation are as follows:

i) the IMF of Miller \& Scalo \shortcite{ms79} is used and the
primary mass is generated from the formula of Eggleton, Fitchett
\& Tout \shortcite{egg89}:
\begin{equation}
M_{\rm 1}=$$0.19X\over (1-X)^{0.75}+0.032(1-X)^{1/4}$$
\end{equation}
where $X$ is a random number uniformly distributed between 0 and
1. The mass ranges from 0.8 to $100M_\odot$.

ii) the mass ratio distribution is quite controversial and, for
simplicity, we only consider a constant mass ratio distribution
\cite{maz92}.
\begin{equation}
n(q')=1,  0\le q' \le 1
\end{equation}
where $q'=1/q=M_2/M_1$.

iii) We assume that all stars are members of binary systems and
the distribution of separations  is constant in log$a$ ({\bf
where} $a$ is separation).
\begin{equation}
an(a)=\left\{
\begin{array}{ll}
\alpha_{\rm sep}(a/a_0)^m, &a \le a_0\\
\alpha_{\rm sep}, & a_0<a<a_1\\
\end{array}
   \right.
\end{equation}
where $\alpha =0.070, a_0=10R_{\odot},a_1=5.75\times
10^6R_{\odot}=0.13{\rm pc}$ and $m=1.2$. This distribution gives
an equal number of wide binary systems per logarithmic interval
and 50 per cent of systems with orbital periods less than 100 yr
\footnote{It is an assumed distribution inferred from observations
of specstropic binaries (see a series of papers of Griffin, R. F
in {\it The Observatory}). Indeed, there are many distributions
used in the literature, but they are all flat for wide binaries,
leading to similar binary population results. The important thing
for the currently adopted distribution is that it implies 50 per
cent of stellar systems with orbital periods less than 100 yr. One
can simply multiply the results with a coefficient if another
percentage is assumed. }\cite{han95}.

The rapid binary evolution code developed by Hurley et al.
\shortcite{hur00,hur02} is employed here. In addition to all
aspects of single-star evolution, this code includes many features
of binaries, i.e. mass transfer, mass accretion, common-envelope
evolution, collisions, supernova kicks and angular momentum loss
mechanisms etc.. In particular, circularization and
synchronization of the orbit by tidal interaction are calculated
for convective, radiative and degenerate damping mechanisms. As a
comparison, we adopt three {\bf criteria} obtained {\bf in}
different ways, i.e. from a polytropic model (equation 1), from
the paper of Hurley et al. \shortcite{hur02} and from this paper.

In the paper of Hurley et al. \shortcite{hur02},
\begin{equation}
q_{\rm c}=[1.67-x+2(M_{\rm c}/M)^5]/2.13,
\end{equation}
where $M_{\rm c}$ and $M$ are the core mass and the total mass of
the donor, respectively, $x$ is the exponent of the mass-radius
relation at constant luminosity for giant stars and equals to 0.3.
This criterion comes from the assumption that the adiabatic
mass-radius exponent of a giant star $\zeta_{\rm ad}=\partial{\rm
ln}R/\partial {\rm ln}M$ equals to the Roche lobe mass-radius
exponent $\zeta_{\rm L}$, where  $\zeta_{\rm L}\approx 2.13q-1.67$
for conservative mass transfer (Tout et al. 1997) and $\zeta_{\rm
ad}\approx-x+2(M_{\rm c}/M)^5$, which is fitted from detailed
stellar models.

From the sample, we obtain 303291 binaries which begin RLOF when
the primary is a giant or more evolved star via Hurley's rapid
binary evolution code. In the following {\bf section} we will give
the consequences from different {\bf criteria} of the dynamical
instability for mass transfer from a giant to a MS companion.

\subsection{The results from Hurley's criterion}
Among the 303291 binaries undergoing mass transfer from a giant or
more evolved star to its companion, most of them eventually
experience CE evolution, {\bf leaving} short-period binaries or
mergers of the two components if {\bf the} CE cannot be stripped
away.\footnote{Because the primaries have left the main sequence,
these mergers are different from those from two main-sequence
components. They are giant stars and then have no contribution to
single BSs. } According to {\bf Hurley's criterion}, only 3746
binaries may avoid CE formation and probably show the
characteristics of some strange objects, e.g. symbiotic stars, BSs
etc. during or after RLOF. Here we are concerned {\bf with} the
outcomes of BSs.

Most of the 3746 binaries which avoid CE formation begin RLOF when
the primaries are on the giant branches, i.e. on the first giant
branch(FGB) for 17 binaries, on the early asymptotic giant branch
(EAGB, after central He burning but before the first thermal
pulse) for 415 binaries, and on the thermally pulsing AGB (TPAGB)
for 3295 binaries. The simulation shows that, 3139 binaries will
pass through BS phase during their lives and 2208 of them have
undergone dynamically stable RLOF before they become
BSs\footnote{We included wind accretion in the code for the reason
presented in section 6, so some BSs can be formed without RLOF.
The simulation here indicates that wind accretion is important for
BS formation, as the secondaries in some binaries cannot be larger
than the turnoff {\bf mass without wind accretion even after RLOF
.}}. This means that RLOF is an important process to increase the
secondary's mass to be larger than the turnoff of a cluster. All
the 2208 binaries have long orbital periods (greater than 1000 d)
and their mass ratios are less than unity at the onset of RLOF,
indicating that the primaries have lost most of their envelopes
prior to RLOF (e.g 2164 binaries have a mass donor being on
TPAGB). As a consequence, the BSs from this way also have long
orbital periods (greater than 1600 d).

\begin{figure}
\centerline{\psfig{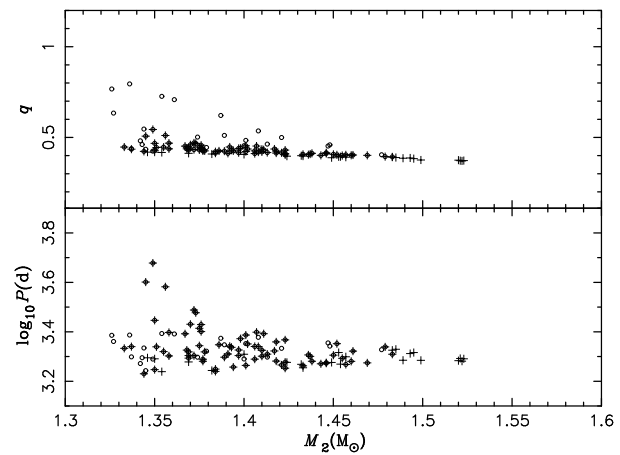}}
\caption{ Mass ratio as well as orbital period versus the mass of
the secondary for 96 binaries which fulfill $t_{\rm 1} \le 4 Gyr <
t_{\rm 2}$, where $t_{\rm 1}$ and $t_{\rm 2}$ are the ages at the
onset and at the end of BSs, respectively. The circles are for the
case at $t_{\rm 1}$ and the pluses are for the case at $t_{\rm
2}$. } \label{h2}
\end{figure}

At the age of 4 Gyr, we obtain 96 BSs. Fig. \ref{h2} presents the
distribution of the orbital period as well as the mass ratio
versus the mass for the 96 BSs. In the figure, we see that, though
the mass ratio at the onset of BS phase is a little different, all
the BSs have similar mass ratios (around 0.4) when they leave the
main sequence. {\bf We explain this as follows. BS formation from
mass transfer between giants and MS companion has some
constraints. For example, RLOF should be dynamically stable. For
$q_{\rm c}$ adopted here, the primaries should be close to or on
the TPAGB before 4 Gyr. Meanwhile, the secondary should be stay on
the MS at 4 Gyr (the MS timescale of the secondary after RLOF is
only in order of $10^8$ yr). So the binaries which may produce BSs
in this cluster are very similar initially, and the products after
RLOF are also similar. The difference of $q$ at the onset of BSs
comes from the fact that RLOF has not terminated in these binaries
at that time.} Although RLOF still exists in some BSs, which
results in the increase of their masses, the orbital period {\bf
do not show} obvious changes during the whole BS phase. All the
BSs are near the turnoff of M67, indicating that this evolution
channel, i.e mass transfer from a giant star to a main sequence
companion(MS), can only account for the BSs in this region.

\subsection{The results from the polytropic model's criterion }
Using the same binary sample, we examined the evolutionary
consequence from the criterion of the polytropic model (see
equation (1) in section 1). As discussed in the paper of Hurley et
al. (2002), the value of $q_{\rm c}$ from the polytropic model is
obviously larger than that of Hurley et al. (2002) with increasing
core mass. For example, it is a factor of 2 larger at $M_{\rm
c}/M_{\rm 1}=0.6$. So more binaries may avoid dynamically unstable
RLOF in the criterion of the polytropic model than those in the
criterion of Hurley et al. (2002).

From equation 1, 8522 binaries avoid CE formation and 5919 pass
through {\bf the} BS phase during their lives. {\bf Among of} the
5919 binaries, 4287 have undergone or are just undergoing
dynamically stable RLOF before the secondaries become BSs. At the
age of 4 Gyr, we obtained 175 long-orbital period BSs (greater
than 1600 d) . The distribution of the orbital period as well as
the mass ratio versus the mass for the 175 BSs are similar to
those from the criterion of Hurley et al. (2002).

\subsection{The results from this paper's criterion}
Now we study the outcomes from the criterion in this paper. {\bf
Different from the two criteria above, we only evolve binaries to
the onset of RLOF using Hurley's code and obtain the following
evolutions from assumptions below.}

When we determine whether mass transfer leads to a BS formation or
not in a certain cluster, two characteristic parameters of the
secondary, i.e. the mass $M_{\rm a}$ and the MS lifetimes $t_{\rm
MS}$, are critical. The former is relevant to the initial mass of
the secondary and the mass transfer efficiency $\beta$, while the
latter is determined by the central H fraction and the total mass
of the secondary after accretion. Meanwhile, the ages {\bf at the
onset and termination} of RLOF are also important. For example,
RLOF should start before or at the cluster age $t_{\rm cluster}$
to ensure that the secondary may accrete some matter, but it
cannot stop much earlier than $t_{\rm cluster}$, or the secondary
has likely left the main sequence. Therefore, only the binaries,
which complete mass transfer between $t_{\rm cluster}-t_{\rm MS}$
and $t_{\rm cluster}$ as well as have secondary's mass $M_{\rm
a}>M_{\rm to}$ during or after accretion, will contribute to BSs
in a cluster. In fact, the mass transfer process from a giant star
to its companion is on a very short time-scale (about $10^6$ yr)
in comparison to the MS time of the secondary after accretion, and
then can be ignored. The amount of matter lost from the mass donor
during RLOF and the mass transfer efficiency $\beta$ finally
determine how much matter is accreted by the secondary. From
section 4, we see that the final mass of the mass donor after RLOF
is a little larger (about $0.1M_\odot$) than the core mass at the
onset of RLOF. Meanwhile, from the studies of the initial-final
mass relation \cite{meng07}, the final mass of a star after RLOF
or the super wind is around the core mass at the onset of RLOF or
before super wind. The maximum difference induced by thermal
pulses and mass loss is less than $0.4M_\odot$, so we choose a
range of the final mass of the mass donor from $M_{\rm c}$ to
$M_{\rm c}+0.4M_\odot$ to examine the characteristics of the
secondaries and of the binary systems.

\begin{figure}
\centerline{\psfig{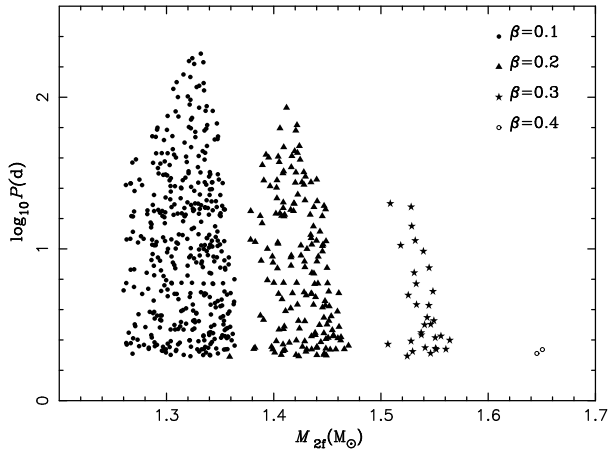}}
\caption{Orbital period versus mass for the BSs in M67 resulting
from FGB stars transferring material to main-sequence companions.
The final mass of the primary is assumed to be $M_{\rm
c}+0.2M_\odot$, where $M_{\rm c}$ is the core mass of the mass
donor at the onset of RLOF. See the text for the details.}
\label{m2p3}
\end{figure}

\begin{figure}
\centerline{\psfig{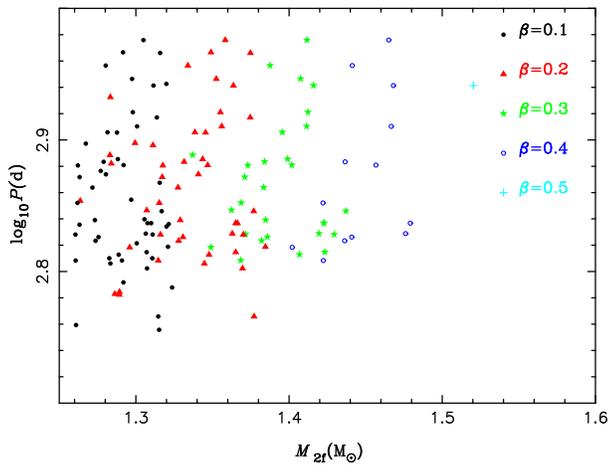}}
\caption{Similar to Fig. \ref{m2p3}, but the mass donors are EAGB
stars, i.e. after central He burning but before the first thermal
pulse.} \label{m2p5}
\end{figure}

\begin{figure}
\centerline{\psfig{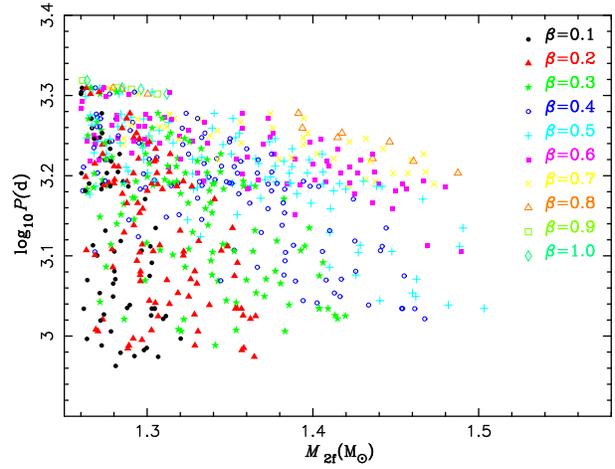}}
\caption{Similar to Fig. \ref{m2p3}, but the mass donors are TPAGB
stars.} \label{m2p6}
\end{figure}

For M67 ($t_{\rm cluster}=4 \times 10^9$ yr), we obtained
different possible candidates from various $\beta$ for $t_{\rm
MS}=1 \times 10^9$ yr and $t_{\rm MS}=5 \times 10^8$ yr,
respectively\footnote{The MS time-scales here are referred to
Table 5{\bf -- it is about $8\times10^8$ yr from the onset of RLOF
to the termination of the secondary on the main sequence.} From
Fig. 17, we see that the maximum mass ratio at the onset of RLOF
is less than 1.2 for binaries resulting in BSs, which means that
the secondary is only slightly less than the turn-off, and then
the MS time-scale of the secondary after accretion will not be
very long. Meanwhile, it is likely longer than $5\times 10^8$ yr
from Table 5.}. The final orbital period of the candidates is
estimated based on the assumption that the lost matter carries off
the specific angular momentum as the accretor. Tables \ref{4} to
\ref{6} present the BS numbers for different mass donors, i.e. the
mass donors are FGB stars (FGB+MS), EAGB stars (EAGB+MS) and TPAGB
stars (TPAGB+MS), respectively. The corresponding distributions of
the orbital period versus the final mass for various $\beta$ are
shown in Figs. \ref{m2p3} to \ref{m2p6}.

We have not obtained BSs from an FGB star transferring matter to a
MS companion in M67 from the {\bf criteria} of Hurley et al.
(2002) and of the polytropic model, since the mass donor has not
lost much mass by wind prior to RLOF and the core mass is not very
large. {\bf Both of these facts make the condition $q<q_{\rm c}$
difficult to fulfill} (see equations (1) and (14)). However, from
the criterion of this paper, several BSs will be formed in M67 if
$\beta \le 0.4$. From Table \ref{4}, we see that $\beta$ strongly
affects the contribution to BSs from this evolutionary channel. A
small $\beta$ contributes more BSs because RLOF is stabilized more
easily. The mass of the secondary in this case is generally large
enough to be larger than the turnoff after accretion even as
$\beta \le 0.1$.

\begin{table}
 \begin{minipage}{85mm}
 \caption{The BS numbers in M67 obtained from FGB stars transferring matter to MS companions
 for various $\beta$. The MS timescale of the secondary after RLOF, $t_{\rm MS}$, is set to
 $1 \times 10^9$ yr (from the 2nd to 5th colomuns) and $5 \times 10^8$ yr (from the 6th to 9th colomuns), respectively.
The final masses of the primary are $M_{\rm 1f}^1=M_{\rm c}$,
$M_{\rm 1f}^2=M_{\rm c}+0.1M_\odot$,$M_{\rm 1f}^3=M_{\rm
c}+0.2M_\odot$ and $M_{\rm 1f}^4=M_{\rm c}+0.4M_\odot$, where
$M_{\rm c}$ is the core mass of the primary at the onset of RLOF.}
 \label{tab4}
   \begin{tabular}{ccccccccc}
\hline
$\beta$&$M_{\rm 1f}^1$&$M_{\rm 1f}^2$&$M_{\rm 1f}^3$&$M_{\rm 1f}^4$&$M_{\rm 1f}^1$&$M_{\rm 1f}^2$&$M_{\rm 1f}^3$&$M_{\rm 1f}^4$\\
\hline
0.1&414&412&406&371&319&317&311&276\\
0.2&169&169&169&169&150&150&150&150\\
0.3&31&31&31&31&30&30&30&30\\
0.4&2&2&2&2&2&2&2&2\\
\hline \label{4}
\end{tabular}
\end{minipage}
\end{table}

\begin{table}
 \begin{minipage}{85mm}
 \caption{The BS numbers in M67 obtained from EAGB stars transferring matter to MS companions
 for various $\beta$. }
 \label{tab5}
   \begin{tabular}{ccccccccc}
\hline
$\beta$&$M_{\rm 1f}^1$&$M_{\rm 1f}^2$&$M_{\rm 1f}^3$&$M_{\rm 1f}^4$&$M_{\rm 1f}^1$&$M_{\rm 1f}^2$&$M_{\rm 1f}^3$&$M_{\rm 1f}^4$\\
\hline
0.1&58&56&52&38&32&30&28&18\\
0.2&48&47&46&37&41&40&39&30\\
0.3&29&29&29&29&28&28&28&28\\
0.4&13&13&13&13&13&13&13&13\\
0.5&1&1&1&1&1&1&1&1\\
\hline \label{5}
\end{tabular}
\end{minipage}
\end{table}

\begin{table}
 \begin{minipage}{85mm}
 \caption{The BS numbers in M67 obtained from TPAGB stars transferring matter to MS companions
 for various $\beta$. }
 \label{tab6}
   \begin{tabular}{ccccccccc}
\hline
$\beta$&$M_{\rm 1f}^1$&$M_{\rm 1f}^2$&$M_{\rm 1f}^3$&$M_{\rm 1f}^4$&$M_{\rm 1f}^1$&$M_{\rm 1f}^2$&$M_{\rm 1f}^3$&$M_{\rm 1f}^4$\\
\hline
0.1&97&80&63&29&41&33&23&9\\
0.2&150&122&96&50&75&60&47&21\\
0.3&175&142&112&61&98&73&57&32\\
0.4&175&146&105&58&104&85&58&32\\
0.5&157&132&97&45&96&79&57&24\\
0.6&111&100&81&31&66&59&47&16\\
0.7&68&56&42&15&43&37&27&10\\
0.8&39&26&11&4&24&17&7&3\\
0.9&32&20&4&0&21&14&4&0\\
1.0&33&20&4&0&22&14&4&0\\
 \hline \label{6}
\end{tabular}
\end{minipage}
\end{table}

\begin{figure}
\centerline{\psfig{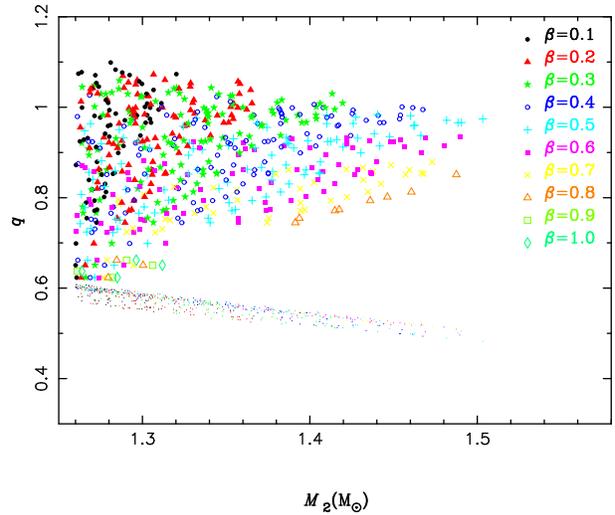}}
\caption{Mass ratio versus the mass for the BSs resulting from
TPAGB stars transferring matter to MS companions for various
$\beta$ when the final mass of the primary is $M_{\rm 1f}=M_{\rm
c}+0.2M_\odot$. The small dots are at the end of RLOF.}
\label{m2q6}
\end{figure}

The contribution to BSs from EAGB+MS, in comparison to that of
TPAGB+MS, is very small, although the value of $q_{\rm c}$ for
EAGB+MS is likely larger than that of TPAGB+MS because of the
small radii at the onset of RLOF. This is relevant to the mass
loss by wind prior to RLOF during the two phases. The stellar wind
is generally much less in EAGB than that in TPAGB. Our study
indicates that the mass loss prior to RLOF during EAGB phase is
usually not enough to make $q<q_{\rm c}$, while {\bf a lot of}
matter in the envelope of the primary has left the system by wind
prior to RLOF for binaries of TPAGB+MS. So {\bf for systems of
TPAGB+MS, the mass transfer is easily stabilized and leads to} the
formation of long-orbital period binaries after RLOF.

The influence of $\beta$ on the BS formation from TPAGB+MS
binaries is non-monotonic,i.e. the BS number first increases, and
then decreases with the increase of $\beta$ (see Table \ref{6}).
This non-monotonic phenomenon is induced by the criterion of
$q_{\rm c}$ and the mass of the MS companion after accretion. With
the increase of $\beta$, $q_{\rm c}$ becomes smaller and smaller,
which leads  $q<q_{\rm c}$ to be fulfilled {\bf more unlikely},
but the MS companion may accrete more material from the primary to
increase the mass to be larger than the turn-off of the cluster.
The results of Table \ref{6} indicate that, when $\beta \le 0.4$,
the mass of the MS companion is the main factor to determine the
formation of BS from this way, but the dynamical instability is
the crucial {\bf factor} when $\beta > 0.4$.

From Fig. \ref{m2p3} we see that, for the BSs from mass transfer
between an FGB star and a MS companion, their masses increase and
the range of the orbital period becomes narrower (mainly some long
orbital period BSs disappear) with the increasing $\beta$, since a
large $\beta$ means more matter accreted by the secondary (leading
to a larger final mass) and less angular momentum lost from the
system (resulting in a shorter orbital period). Furthermore, a
long-orbital period indicates a larger stellar radius at the onset
of RLOF, and then a smaller $q_{\rm c}$ as shown in section 3.
Therefore, with the increase of $\beta$, the RLOF in relatively
long-orbital-period binaries firstly changes from dynamically
stable to dynamically unstable, {\bf and} directly leads to the
disappearance of long-orbital-period BSs. The orbital period of
the BSs from this {\bf channel} has a wide range, i.e. from
several days to hundreds of days for $\beta=0.1$. However, BSs
with $P \ge 100$ days can only be produced when $\beta \le 0.2$
and their masses are near the turnoff because only a very small
fraction of the lost matter from the primary is accreted by the
secondary. All the possible candidates contributing to BSs in M67
from FGB+MS are low-mass binaries, since the secondaries from
intermediate-mass FGB+MS binaries have left the main sequence at 4
Gyr. For a $2M_\odot$ star, the minimum mass of the secondary is
about $1.6M_\odot$ for $\beta=0.0$ and $M_{\rm c}=0.25M_\odot$.
The stars with this mass have left the main sequence in the old
open cluster.

Different from the case of FGB+MS, the BSs resulting from EAGB+MS
and TPAGB+MS have similar orbital periods for various $\beta$ (see
Figs. \ref{m2p5} and \ref{m2p6}). Meanwhile, the mass difference
of the BSs from different $\beta$ is also not as obvious as that
from FGB+MS binaries. {\bf In the cases of EAGB+MS and TPAGB+MS,
some BSs have masses similar to those from FGB+MS, and some have
lower masses, when $\beta$ is larger than 0.1. The lower mass BSs
mix with those from $\beta$ with lower than the given value.
Stellar wind may account for this consequence. When the primary
evolves from FGB to EAGB, then to TPAGB, its mass becomes less and
less due to stellar wind. Correspondingly, the minimum mass of the
MS companion for stable mass transfer also becomes smaller. The
companions may then have lower masses after RLOF with evolution.
As a result, the masses of blue stragglers in Figs. 15 and 16
extend to lower values than those in Fig.14. }

From the {\bf criteria} of {\bf the} polytropic model and Hurley
et al. \shortcite{hur02}, only long-orbital period BSs (with
orbital period greater than 1600 d) may be produced from giants
transferring matter to the MS companions. Since case A and early
case B (the mass donor is during Hertzsprung gap at the onset of
RLOF) mass transfer are only responsible for some short- and
mid-period BSs (up to 100 d, Chen \& Han, 2004), there seems to be
a period gap of hundreds days from the two {\bf criteria} above.
However, this period gap will not appear from the criterion of
this paper. The BSs from different giant binaries cover the period
range from several days to thousands of days (see Figs. \ref{m2p3}
to \ref{m2p6}).

As an example, we show the mass ratio versus the mass of BSs for
the case of TPAGB+MS binaries at the onset and at the termination
of RLOF in Fig. \ref{m2q6}. From the figure, we see that the mass
ratio at the onset of RLOF is located in a triangle region, where
the upper boundary is determined by the dynamical instability and
the lower boundary is dertermined by the mass of the accretor
after the RLOF.

We have not included example 3 in section 4 in the above study.
Systematic investigation for this case is difficult at present,
since we have little knowledge {\bf of} the conditions for
binaries to avoid CE during initially dynamically unstable RLOF.
The conditions are possibly relevant to the parameters such as the
mass (and the core mass) of the primary, the mass ratio of the
components, the orbital period etc.. Furthermore, it may also be
affected by the detailed process of mass transfer. The study of
the conditions is beyond the scope of {\bf this} paper. If case 3
is a real case, only a very small fraction (less than 1 per cent)
from those with dynamically unstable RLOF is enough to produce BSs
in M67. The products here have much longer main-sequence lives in
comparison to those from dynamically stable RLOF, since the
secondaries in general have much lower mass due to the initially
large mass ratio (greater than $q_{\rm c}$) and are less evolved
before RLOF. From Table \ref{2}, we see that at $t=2.47 {\rm
Gyr}$, a star with $1.5M_\odot$ has already formed and it becomes
a BS when $t > 2.5 {\rm Gyr}$. \footnote{The age $t=2.5 {\rm Gyr}$
is for a cluster with a turnoff of around $1.5M_\odot$.}The
material around the BSs might give us some clues on their parent
stars.

\subsection{Comparison to Observations}
Several works focus on the BSs in M67 from observations
\cite{ml92,al95,ss03}. In the new catalogue of BSs in open
clusters \cite{al07}, there are 30 BSs in this old open cluster.
Since {\bf there is} no orbital information for these BSs in
Ahumada \& Lapasset \shortcite{al07}, the main observational data
we adopted here are from some earlier studies.

Milone \& Latham \shortcite{ml92} reported the radial velocities
for 13 BSs in the open cluster M67 according to observations {\bf
over} about 9 years. Three of the 13 BSs rotate too rapidly to
allow reliable velocity determinations \cite{lm96}. Among the ten
BSs left, only one (F190) shows a short orbital period, about 4.2
d, while five {\bf have} long-orbital periods in the range from
800 to 5000 d. The other four BSs are considered as single stars
by some studies (e.g Hurley et al., 2001). Three of the five
long-orbital period BSs have obvious orbital eccentricities and
the other two are in near-circular orbits. Sandquist \& Shetrone
\shortcite{ss03} presented an analysis of the time series
photometry of M67 for W UMa systems, BSs and related objects.
There are 24 possible BSs observed in their study and most of them
show no variation in {\bf their} light curves. Two BSs, S968 and
S1263, which {\bf were} considered as single stars before, are
possible variables in the study of Sandquist \& Shetrone
\shortcite{ss03}.

The BSs resulting from the criterion of Hurley et al.
\shortcite{hur02} have orbital periods beyond 1600 d (${\rm log}P
>3.2$ as seen in Fig. \ref{h2}). However, four of the five
observed long-orbital period BSs have orbital periods less than
this value. The simulation from the polytropic model {\bf gives} a
similar result. The BSs from the criterion of this paper may well
cover the orbital period range, but it seems that none of the five
BSs is from the mass transfer between an FGB and a MS, since the
smallest orbital period is 846 d (${\rm log}P =2.93$), which is
far greater than the maximum orbital period resulting from the
FGB+MS binaries, as seen in Fig. \ref{m2p3}. The BSs with mid
orbital period (i.e. from tens of days to hundreds of days) are
possibly from {\bf FGBs transferring matter to their MS
companions}, but it needs more observational evidence to comfirm
the existence of these BSs.

The orbital eccentricity is a puzzle from mass transfer, since the
orbit should be circularized by tidal interaction of the two
components of a binary, which is just undergoing, or after, RLOF.
Some studies considered that dynamical collision is necessary to
explain the observed orbital eccentricities. Recently, Marinovic
et al. \shortcite{mar07} found that, due to the enhanced mass loss
of the AGB component at orbital phases closer to the periastron,
the net eccentricity growth rate in one orbit is comparable to the
rate of tidal circularisation in many cases. They reproduced the
orbital period and eccentricity of {\bf the} Sirius system with
this eccentricity enhancing mechanism. Their study provides an
explanation for the eccentricities of long-orbital period BSs
without dynamical collision.

\section{discussions and conclusions}
Our study shows that the critical mass ratio, $q_{\rm c}$, for
dynamically stable mass transfer between a giant star and a MS
companion depends on the stellar radius at the onset of mass
transfer. For {\bf any given} mass, a more evolved star is less
stable for RLOF, since the evolutionary timescale for a more
evolved star is shorter and hence the mass transfer rate is higher
than that {\bf for less-evolved ones}. The results including
stellar wind \cite{han02} are consistent with the tendency in this
paper, except for {\bf cases where $M_{\rm 1i}=0.8M_\odot$}, which
have a larger $f_{\rm c}$ and a larger $q_{\rm c}$ in comparison
to those of $M_{\rm 1i} \ge 1.0M_\odot$. The behaviour of the
cases of $M_{\rm 1i}=0.8M_\odot$ here is similar to the
consequences of a polytropic model. Meanwhile, from Table 3 in the
paper of Han et al. \shortcite{han02}, we see that, {\it at a
similar core mass}, {\bf binaries} with a low primary mass (or a
large $f_{\rm c}$ ){\bf have larger $q_{\rm c}$, also} consistent
with equation(1) except for the binaries with $M_{\rm
1i}=1.9M_\odot$. Our results for $\beta =0$ in this paper {\bf
match those} of Han et al. \shortcite{han02} (see the upper panel
of Fig.\ref{low-fit}). Due to the various core masses, however,
the dependance of $q_{\rm c}$ on $f_{\rm c}$ is completely
scattered as shown in the upper panel of Fig.\ref{low-fit}. So, it
is very likely that both $M_{\rm c}$ and $f_{\rm c}$ have
influences on $q_{\rm c}$, but the effect is different for
binaries {\bf of different} masses, i.e. $M_{\rm c}$ dominates the
case {\bf where} the primary's mass {\bf is} larger than
$1M_\odot$ while $f_{\rm c}$ is critical for the less {\bf
massive} ones. The thickness of the envelope might be an important
cause here and the transition is possibly a gradual process. The
non-monotonic behaviour of $M_{\rm 1i}=1.9M_\odot$ results from
the low degeneracy degree of the core in {\bf primaries} with this
mass.

{\bf From the criterion in this paper, we obtained some BSs with
intermediate orbital period when $\beta$ is less than 0.4.
However, there is no one reported at present located in the
orbital period range from FGB+MS binaries, as we compare our
results to observations. There might be two factors for this
contradiction. One is the fact that the mass of the BSs formed
from such a low $\beta$ is very close to the turnoff of M67,
making them hard to distinguish from normal stars around the
turnoff (as seen in Figure 14). The other might be that the value
of $\beta$ is substantially larger than 0.1 during mass transfer
between FGB stars and MS companions in a real case, so few BSs are
produced in this way. Orbital determinations (in the future) might
provide some constraints on the value of $\beta$. For example,
since the products of FGB+MS have a wide range of orbital periods,
from several days to hundreds of days, there would be some BSs
with intermediate orbital periods (if $\beta$ is low enough),
which are possibly absent from the criterion of Hurley et al.
\shortcite{hur02}.}

The possibility that a main sequence star becomes a BS via wind
accretion was first suggested by Williams \shortcite{wil64}. This
idea lacked attention for a long time, since an isolated star can
hardly accrete enough matter to become a BS. However it is likely
different if the main sequence star is bound in a binary system
where the primary is undergoing a large mass loss. In general, the
mass loss rate of stellar wind is in the range of $10^{-2}M_\odot
{\rm yr}^{-1}$ to $10^{-6}M_\odot {\rm yr}^{-1}$ as a star
approaches {\bf the tip of the AGB}, and some fraction of the lost
material (up to 10 per cent, Tom et al. 1996) {\bf is} probably
accreted by its companion. This means that the secondary may
significantly increase its mass when the primary undergoes a large
mass loss at the tip of {\bf the} AGB. Although the mass increase
is likely not enough to produce a BS, the secondary still has
chances to obtain matter from the primary in the following stable
RLOF if it happens. So BSs with very long orbital periods (greater
than 1000 days) are likely the consequences of both RLOF and wind
accretion.

In {\bf this} paper we present an example of a binary which avoids
CE formation from initially dynamically unstable RLOF and
eventually evolves to a BS with a long orbital period. {\bf We}
only assume that {\bf a} CE has not formed in the initially
dynamically unstable RLOF, but no critical conditions are shown.
How to discriminate whether {\bf a} CE has developed or not during
this phase is unclear, as mentioned in section 5. Many parameters,
such as the mass (and the core mass) of the primary, the mass
ratio of the components, the orbital period {\bf etc}. are
relevant to this condition, which {\bf is} also affected by the
detailed process of mass transfer. The study {\bf of} this will be
{\bf complex} and difficult, while interesting. As long as the
possibility of this case exists, only a very {\bf small fraction
of systems} with dynamically unstable RLOF may provide an
important contribution to BSs.

Mass transfer efficiency, $\beta$, is an important parameter. Both
the criterion of dynamical instability, $q_{\rm c}$, and the final
mass of the accretor, $M_{\rm a}$ ({\bf which are the two critical
characters to determine the formation of a BS}) are relevant to
this parameter. From our simulation for M67, the peak of the
contribution from FGB+MS binaries is {\bf probably} about
$\beta=0.1$, with which the BS mass is near $1.3M_\odot$. With the
increase of $\beta$, the orbital period decreases and the BS mass
increases, as was explained in section 5. However it is unclear
{\bf what the value of $\beta$ is} in binaries. In general,
$\beta$ will be very small for RLOF in binaries with a giant mass
donor and a compact companion \cite{han02}. So $M_{\rm a}$ will
not be very large, {\bf similar to the turnoff mass.}

The detailed evolution calculations in the paper show that $q_{\rm
c}$ decreases with the stellar radii of the primaries at the onset
of RLOF, except for cases at or near the base of {\bf the} giant
branch, where the core is not very degenerate and the envelope is
not yet fully convective. Non-conservative assumptions will
strongly affect $q_{\rm c}$ while stellar wind before mass
transfer has little influence on it. To conveniently use the
result we give a fit of $q_{\rm c}$ as a function of the stellar
radius of the primary at the onset of RLOF, and of the mass
transfer efficiency during RLOF. Theoretically, dynamically stable
mass transfer occurs once the mass ratio is less than the critical
{\bf value}. However, it is delayed in real binaries. Usually the
stable mass transfer occurs after the reversion of the mass ratio.

The Monte Carlo simulations show that some binaries with the mass
donor {\bf on} the first giant branch, which have no contributions
to the blue stragglers from the earlier {\bf criteria}, will
contribute to this population with the criterion obtained in this
paper. Meanwhile, from our criterion, the blue stragglers
resulting from the mass transfer between an AGB star and a MS
companion may be more numerous and have a wider range of orbital
periods than those {\bf formed from previous criteria}. Although
the result from our criterion may well cover the observed orbital
period range, it seems that none of the five observed long-orbital
period BSs is from the mass transfer between an FGB and a MS
companion.

\section{ACKNOWLEDGMENTS}
{\bf The authors thank the referee for his useful suggestions on
this paper and R. S. Pokorny for his improvement of the English in
the manuscript.} This work is supported by the Chinese National
Science Foundation (Grant Nos. 10603013 and 10433030,10521001 and
2007CB815406) and the Chinese Academy of Sciences (Grant No.
O6YQ011001).

\end{document}